\documentclass[fleqn,usenatbib]{rasti}

\usepackage{newtxtext,newtxmath}

\usepackage[T1]{fontenc}
\usepackage{listings}
\usepackage[dvipsnames,table,svgnames]{xcolor}
\usepackage{pgfplots}
\usetikzlibrary{pgfplots.groupplots,intersections}
\usepgfplotslibrary{fillbetween}
\pgfplotsset{compat=1.18}
\DeclareRobustCommand{\VAN}[3]{#2}
\let\VANthebibliography\thebibliography
\def\thebibliography{\DeclareRobustCommand{\VAN}[3]{##3}\VANthebibliography}
\lstdefinelanguage{GLSL}%
{%
	morekeywords={%
		false,FALSE,NULL,true,TRUE,%
		__LINE__,__FILE__,__VERSION__,GL_core_profile,GL_es_profile,GL_compatibility_profile,%
		precision,highp,mediump,lowp,%
		break,case,continue,default,discard,do,else,for,if,return,switch,while,%
		void,bool,int,uint,float,double,vec2,vec3,vec4,dvec2,dvec3,dvec4,bvec2,bvec3,bvec4,ivec2,ivec3,ivec4,uvec2,uvec3,uvec4,mat2,mat3,mat4,mat2x2,mat2x3,mat2x4,mat3x2,mat3x3,mat3x4,mat4x2,mat4x3,mat4x4,dmat2,dmat3,dmat4,dmat2x2,dmat2x3,dmat2x4,dmat3x2,dmat3x3,dmat3x4,dmat4x2,dmat4x3,dmat4x4,sampler1D,sampler2D,sampler3D,image1D,image2D,image3D,samplerCube,imageCube,sampler2DRect,image2DRect,sampler1DArray,sampler2DArray,image1DArray,image2DArray,samplerBuffer,imageBuffer,sampler2DMS,image2DMS,sampler2DMSArray,image2DMSArray,samplerCubeArray,imageCubeArray,sampler1DShadow,sampler2DShadow,sampler2DRectShadow,sampler1DArrayShadow,sampler2DArrayShadow,samplerCubeShadow,samplerCubeArrayShadow,isampler1D,isampler2D,isampler3D,iimage1D,iimage2D,iimage3D,isamplerCube,iimageCube,isampler2DRect,iimage2DRect,isampler1DArray,isampler2DArray,iimage1DArray,iimage2DArray,isamplerBuffer,iimageBuffer,isampler2DMS,iimage2DMS,isampler2DMSArray,iimage2DMSArray,isamplerCubeArray,iimageCubeArray,atomic_uint,usampler1D,usampler2D,usampler3D,uimage1D,uimage2D,uimage3D,usamplerCube,uimageCube,usampler2DRect,uimage2DRect,usampler1DArray,usampler2DArray,uimage1DArray,uimage2DArray,usamplerBuffer,uimageBuffer,usampler2DMS,uimage2DMS,usampler2DMSArray,uimage2DMSArray,usamplerCubeArray,uimageCubeArray,struct,%
		gl_BackColor,gl_BackLightModelProduct,gl_BackLightProduct,gl_BackMaterial,gl_BackSecondaryColor,gl_ClipDistance,gl_ClipPlane,gl_ClipVertex,gl_Color,gl_DepthRange,gl_DepthRangeParameters,gl_EyePlaneQ,gl_EyePlaneR,gl_EyePlaneS,gl_EyePlaneT,gl_Fog,gl_FogCoord,gl_FogFragCoord,gl_FogParameters,gl_FragColor,gl_FragCoord,gl_FragData,gl_FragDepth,gl_FrontColor,gl_FrontFacing,gl_FrontLightModelProduct,gl_FrontLightProduct,gl_FrontMaterial,gl_FrontSecondaryColor,gl_InstanceID,gl_Layer,gl_LightModel,gl_LightModelParameters,gl_LightModelProducts,gl_LightProducts,gl_LightSource,gl_LightSourceParameters,gl_MaterialParameters,gl_ModelViewMatrix,gl_ModelViewMatrixInverse,gl_ModelViewMatrixInverseTranspose,gl_ModelViewMatrixTranspose,gl_ModelViewProjectionMatrix,gl_ModelViewProjectionMatrixInverse,gl_ModelViewProjectionMatrixInverseTranspose,gl_ModelViewProjectionMatrixTranspose,gl_MultiTexCoord0,gl_MultiTexCoord1,gl_MultiTexCoord2,gl_MultiTexCoord3,gl_MultiTexCoord4,gl_MultiTexCoord5,gl_MultiTexCoord6,gl_MultiTexCoord7,gl_Normal,gl_NormalMatrix,gl_NormalScale,gl_ObjectPlaneQ,gl_ObjectPlaneR,gl_ObjectPlaneS,gl_ObjectPlaneT,gl_Point,gl_PointCoord,gl_PointParameters,gl_PointSize,gl_Position,gl_PrimitiveIDIn,gl_ProjectionMatrix,gl_ProjectionMatrixInverse,gl_ProjectionMatrixInverseTranspose,gl_ProjectionMatrixTranspose,gl_SecondaryColor,gl_TexCoord,gl_TextureEnvColor,gl_TextureMatrix,gl_TextureMatrixInverse,gl_TextureMatrixInverseTranspose,gl_TextureMatrixTranspose,gl_Vertex,gl_VertexID,%
		gl_MaxClipPlanes,gl_MaxCombinedTextureImageUnits,gl_MaxDrawBuffers,gl_MaxFragmentUniformComponents,gl_MaxLights,gl_MaxTextureCoords,gl_MaxTextureImageUnits,gl_MaxTextureUnits,gl_MaxVaryingFloats,gl_MaxVertexAttribs,gl_MaxVertexTextureImageUnits,gl_MaxVertexUniformComponents,%
		abs,acos,all,any,asin,atan,ceil,clamp,cos,cross,degrees,dFdx,dFdy,distance,dot,equal,exp,exp2,faceforward,floor,fract,ftransform,fwidth,greaterThan,greaterThanEqual,inversesqrt,length,lessThan,lessThanEqual,log,log2,matrixCompMult,max,min,mix,mod,noise1,noise2,noise3,noise4,normalize,not,notEqual,outerProduct,pow,radians,reflect,refract,shadow1D,shadow1DLod,shadow1DProj,shadow1DProjLod,shadow2D,shadow2DLod,shadow2DProj,shadow2DProjLod,sign,sin,smoothstep,sqrt,step,tan,texture1D,texture1DLod,texture1DProj,texture1DProjLod,texture2D,texture2DLod,texture2DProj,texture2DProjLod,texture3D,texture3DLod,texture3DProj,texture3DProjLod,textureCube,textureCubeLod,transpose,%
		rgb,in,out,uniform,const
	},
	sensitive=true,%
	morecomment=[s]{/*}{*/},%
	morecomment=[l]//,%
	morestring=[b]",%
	morestring=[b]',%
	moredelim=*[directive]\#,%
	moredirectives={define,defined,elif,else,if,ifdef,endif,line,error,ifndef,include,pragma,undef,warning,extension,version}%
}[keywords,comments,strings,directives]%
\usepackage{graphicx}	
\usepackage{amsmath}	
\usepackage[table,svgnames]{xcolor}

\title[All-sky cloud detection]{Nighttime Cloud Detection, Tracking and Prediction with All-Sky Cameras}

\author[S. Buntin et al.]{
Sebastian Buntin,$^{1}$\thanks{E-mail: s.buntin@2019.ljmu.ac.uk}
Chris M. Copperwheat,$^{1}$
Helen E. Jermak$^{1}$
\\
$^{1}$Astrophysics Research Institute, Liverpool John Moores University, IC2, Liverpool Science Park, 146 Brownlow Hill, Liverpool, L3 5RF, United Kingdom \\
}

\date{Accepted XXX. Received YYY; in original form ZZZ}

\pubyear{2024}

\begin{document}
\label{firstpage}
\pagerange{\pageref{firstpage}--\pageref{lastpage}}
\maketitle

\begin{abstract}
This paper presents a novel method for real-time nighttime cloud detection, tracking, and prediction using all-sky cameras, aimed at enhancing the efficiency of ground-based robotic telescopes. Ground-based telescopes are vulnerable to adverse weather conditions, particularly cloud cover, which can lead to the loss of valuable observation time and potential damage to the telescope. Existing methods for cloud detection have limitations in accuracy, particularly under varying illumination conditions such as gibbous moon phases. To address these challenges, we developed an algorithm that uses the temporal incoherence of image sequences from all-sky cameras. The method computes difference images to highlight moving cloud structures, applies Otsu thresholding to generate binary cloud maps, and uses mathematical morphology techniques to reduce noise from bright stars and other artifacts. Segmented cloud regions are then tracked across successive frames, allowing estimation of a velocity vector and enabling short-term predictions of cloud movement. Our approach achieves reliable cloud detection and tracking, providing predictions up to 15 minutes into the future — a capability critical for robotic telescopes that rely on look-ahead scheduling. The system was validated against extensive historical data from the Liverpool Telescope's Skycam A and T systems, achieving a false positive rate of approximately 1\% and a similar false negative rate, depending on cloud thickness and speed. By improving cloud forecasting and observational scheduling, the system offers a valuable tool for enhancing the operational reliability of robotic telescopes.
\end{abstract}




\section{Introduction}
\label{sec:intro}
Ground-based telescopes are susceptible to  environmental factors that can adversely affect their operational integrity and scientific output. Among these environmental factors, precipitation and elevated levels of relative humidity stand out as threats, possessing the potential to inflict harm to the optical components that are fundamental to the telescopes' function and the electronic systems that control their operations. To protect the telescope from these effects, the telescope's enclosure must be closed before the environment changes to these potentially harmful conditions. Robotic telescopes like the Liverpool Telescope (LT, \cite{steele2004}) do not have human nighttime operators who can observe the weather conditions and decide if the telescope can be operates safely or not. The decision must be made by a computer system and ensure safety at all times. This can lead to very conservative strategies where the enclosure is shut in conditions where a human operator would decide otherwise. 

In situations where the sky is partially cloudy and it is still safe for the telescope to operate, the telescope might observe cloudy areas and produce unusable blank frames, wasting valuable observation time, especially if other parts of the sky could have been observed instead. So it is helpful to know before an observation whether the target is visible or covered by clouds. 

The Liverpool Telescope uses a dispatch scheduler, in which the robotic control software selects a new observing group from those available when the current group is concluded. One consequence of this is that a high-science-priority group might not be chosen if it is at a low altitude and rising, since the robotic controller will prefer to wait and observe it at a lower airmass. However, if weather conditions are worsening, this may be an undesirable choice. The New Robotic Telescope (NRT, \cite{copperwheat2015}) is therefore planned to be equipped with a look-ahead scheduler, a scheduling system that plans several observations ahead. This type of scheduler must be aware not only of the current cloud situation but also of the situation in the near future. The resolution of the data from weather prediction models is usually too coarse in the spatial and temporal dimensions for a detailed cloud pattern over the observatory.  

Automated cloud-detection is still an active research field. In recent years, some algorithms proposed for cloud detection, including  \cite{devNighttimeSkyCloud2017}, make use of so called superpixels to detect clouds. The algorithm tries to cluster segments in the image using spatial coherence and then distinguish those segments as either cloud or sky. The algorithm performs well in dark conditions but encounters problems with overexposed areas in the image, occurring especially in the gibbous moon phases. This can result in large sections of the sky around the moon being clustered to one section and labelled as cloud, no matter the true conditions.

Another method, proposed by \cite{adam2017}, tries to solve the problem of finding cloud-covered and open parts in the sky by detecting stars in the night sky and marking regions where no or only very bright stars are detected as cloud-covered. This method works also well on dark nights but has problems in nights during gibbous lunar phases where most stars are not visible in the all-sky image due to over-exposure effects. The authors also use their cloud map as one object and track its movement across the sky. Problems occur again on full moon nights. The moon itself creates a large area in the image due to the overexposure caused by fixed exposure times, lens flares and dust on the lens. This overexposed area is moving with the apparent movement of the night sky and can cover a large part of the image. In this area, star detection is not possible.

More recent and computationally more expensive methods, as shown in \cite{mommert2020}, use neural networks to detect clouds in all-sky images. Although neural networks are an interesting solution, the system needs a lot of accurately annotated training data to provide results comparable to the ones presented in this paper.

In a paper published by \citep{ye2022}, a self-training algorithm using superpixel segmentation and machine learning is proposed for daytime sky images. The algorithm is able to train itself with a few incompletely annotated cloud images, thus reducing the training and annotation amount significantly but relies on the colour difference in the image and is thus not applicable to night-time images.

Cloud tracking is also an active field of research with many applications in the fields of weather forecasting, the energy sector (for estimating the energy produced by photovoltaic systems in smart grids), climate research and also astronomy. The tracking is done using different methods, for example by calculating the optical flow (see \cite{zhang2019}). This approach requires a high image rate (at least 20 all-sky images per minute) to produce usable results and is not able to handle changes in the cloud structure well. Small changes in the cloud structure or a too large time gap between two frames will lead to physically implausible results. 

\citep{peng2015} use several all sky cameras distributed over a large area to generate a full 3D cloud model, including the cloud height. This allows very accurate detection and tracking but is not feasible at the Observatorio del Roque de Los Muchachos (ORM) as the area necessary to distribute the cameras is not available.

For a general overview, \citep{zaher2017} conducted a comparative study of the algorithms available when writing their paper and \citep{arrais2022} published an overview study on cloud tracking methods for short-term forecasting. 

In this paper, a new method based on the temporal information stored in image-sequences taken by the all-sky camera at the Liverpool Telescope is introduced. Section \ref{sec:clouddetection} explains the cloud detection algorithm used to create the cloud maps. Section \ref{sec:cloudmovementprediction} explains how these cloud maps are then used to track clouds in the sky and predict their future location. Section \ref{sec:rd} then describes how the algorithms were tested and explains the results of these tests. The method can be easily adopted for any all-sky camera system, independent of the cameras resolution or location as long as a constant time series of images is produced and a World-Coordinate System (WCS) can be fitted by a tool like Astrometry.net \citep{lang2010}.

\section{Cloud Detection}
\label{sec:clouddetection}
All discussed cloud detection methods make use of one single image and apply methods for detecting clouds. All-sky cameras usually take images with a set frequency. For example, the all-sky camera at the Liverpool Telescope (Skycam A, \cite{mawson2013}) takes images with an exposure time of 30 second once every minute. This results in a regular sampling of the night sky, adding another dimension to the data. With images taken at regular intervals, the temporal incoherence of the cloud structure in those images can be used to detect and track them. As clouds move with a relatively constant velocity in the night sky, the difference of two adjacent images (or even images taken over a span of 3 or more minutes) can be used to detect clouds. 

\subsection{Description of the Algorithm}

Typically, from image to image, the only moving (changing) objects are the apparent movement of the night sky, the movement of the telescope structure and the movement of the clouds. 

We use OpenGL and the GPU’s parallel processing capabilities to accelerate image normalization and difference image calculation.

The algorithm described here uses well-established image processing techniques in a novel way - to detect night-time clouds in monochromatic all-sky images. A lighting preprocessing step, critical for compensating for overexposures caused by moonlight and for normalizing pixel intensities across the image is applied first. Here, we use the FITS header parameters \texttt{BSCALE} and \texttt{BZERO} to linearly transform the raw pixel values to physical intensity values, following the relation: $physical\_value = BSCALE \times pixel\_value + BZERO$. The \texttt{BSCALE} parameter defines the scaling factor applied to each pixel, while \texttt{BZERO} represents an offset (bias) added after scaling. These parameters are commonly used to store integer data in FITS files while preserving floating-point precision.

We then apply a capping parameter \texttt{MAX} to limit the maximum allowed pixel intensity after scaling. This threshold is defined empirically based on typical saturation levels of the camera sensor and is used to compress the dynamic range, preventing bright regions (e.g., the Moon-disk) from dominating the subsequent processing.

This normalization allows subsequent difference imaging and thresholding to work effectively across a wide range of illumination conditions, including gibbous and full moon phases.

On the GPU, the FITS data (in our case with a resolution of 1392x1040 pixels) are loaded as 2D textures into texture memory, and a pixel shader is used to apply the normalization operation across all pixels in parallel. The shader program calculates the scaled and clipped pixel intensity for each pixel and outputs the result as a processed image ready for difference calculation.
\begin{figure}
\centering
  \includegraphics[width=\linewidth]{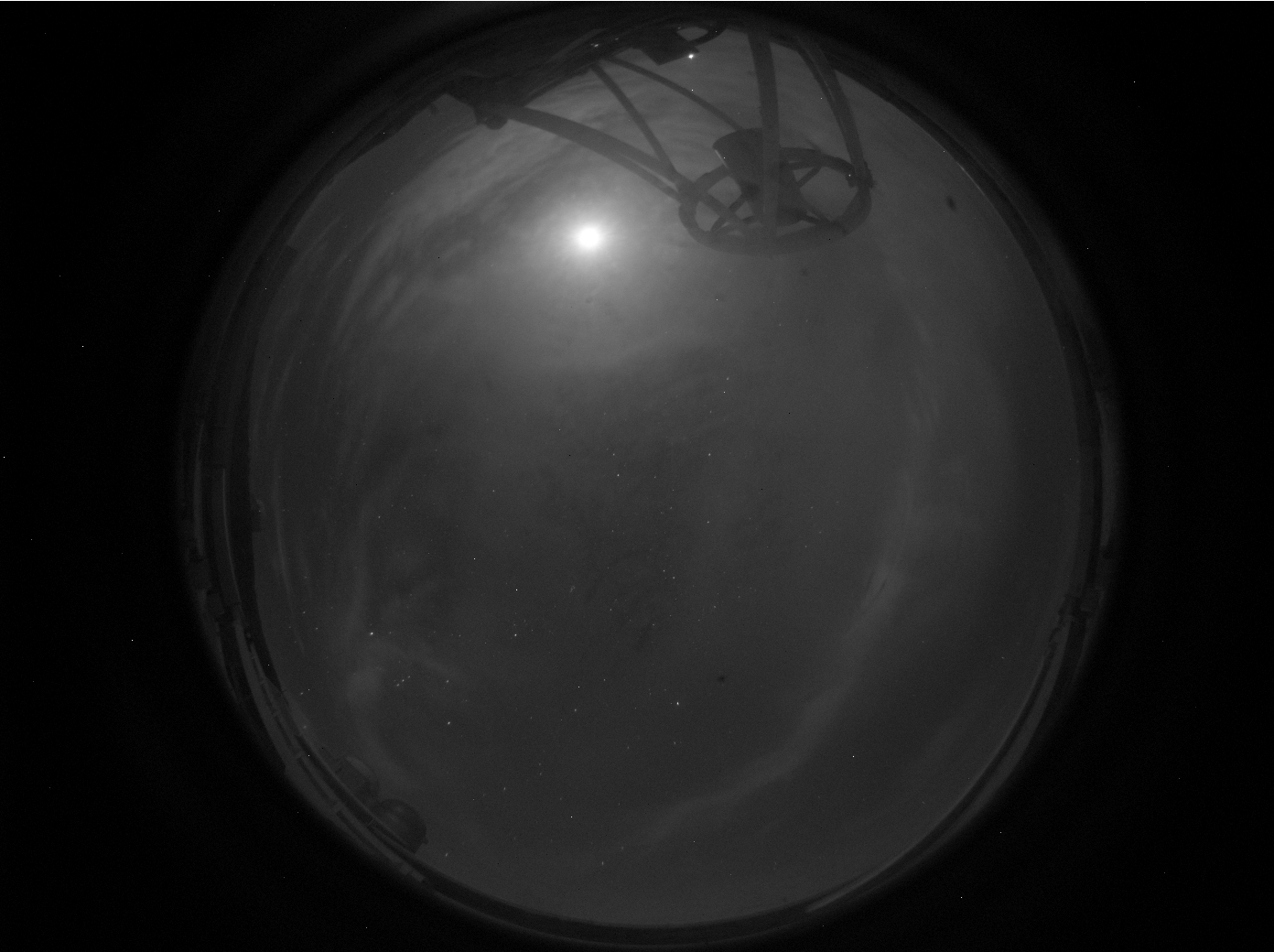}
  \caption{All-sky image taken  using the Skycam a Camera at the Liverpool Telescope site at 22:45 UT on 2020 Jan 01. Some stars are visible as well as bands of clouds, with the cloud illuminated by the Moon at the top of the image}
  \label{fig:img1}
\end{figure}

\begin{figure}
\centering
  \includegraphics[width=\linewidth]{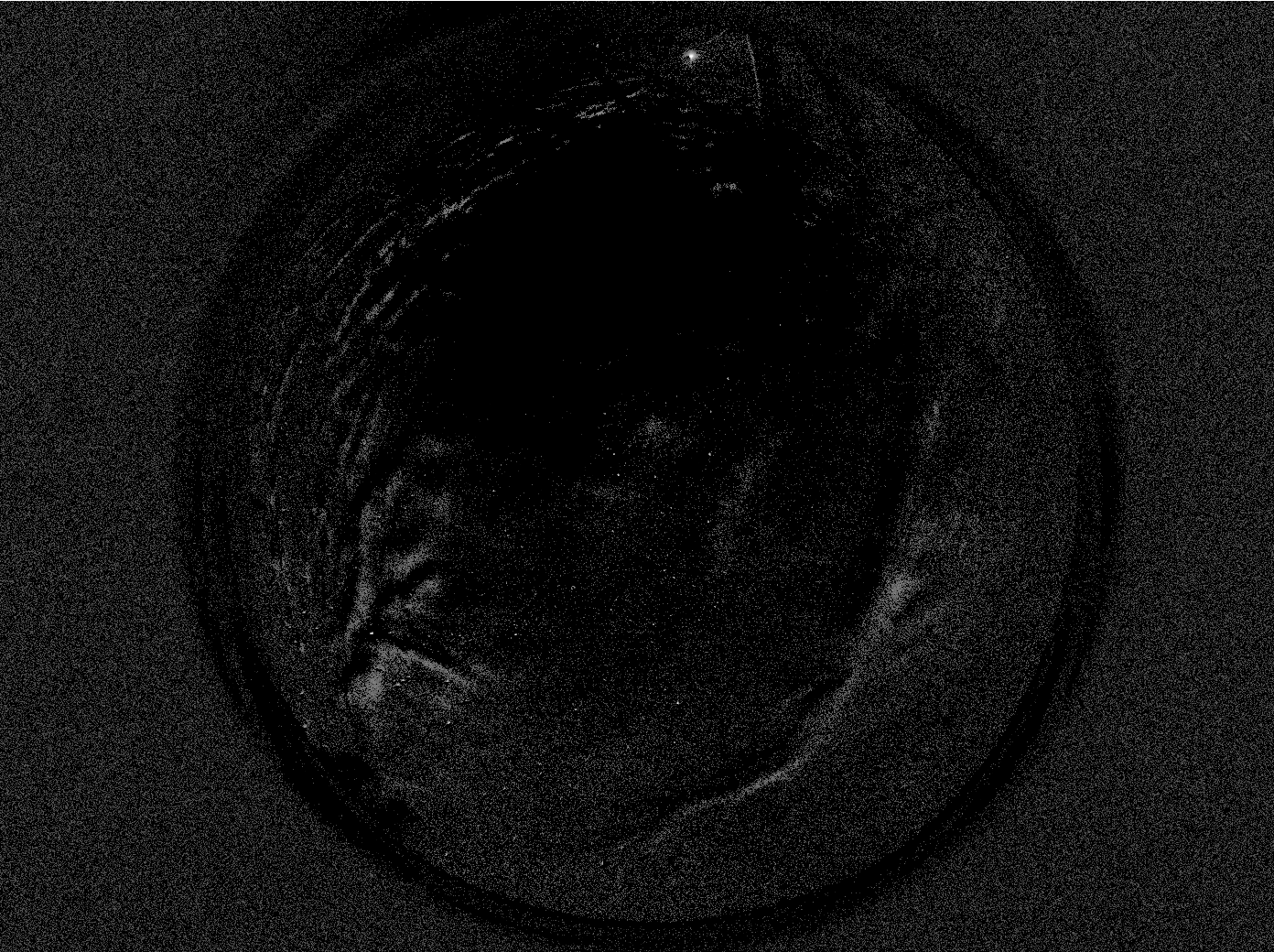}
  \caption{Difference Image obtained by subtracting the image shown in figure \ref{fig:img1} from the image obtained one minute previously. The majority of the structure observed in this image is due to the motion of the cloud between the two exposures}
  \label{fig:diffimg}
\end{figure}

Then, a difference image (see figure \ref{fig:diffimg}) of the current (Figure \ref{fig:img1}) and the previous lighting-corrected all-sky images is taken. This is again done using a shader program. Here, both images (the current one and the `past' image) are loaded as textures for a pixel shader which then calculated the absolute difference of each pixel. 

While the presented implementation leverages GPU hardware for real-time difference image calculation and preprocessing, we note that the processing load (~1 million pixel images at 1-minute cadence) is well within the capabilities of modern multi-core CPUs. The use of GPU acceleration was motivated by our aim to scale towards higher-cadence and higher-resolution cameras especially in the light of the upcoming New Robotic Telescope, as well as to leverage available hardware for parallel processing. However, real-time performance for the given resolution and cadence is achievable on a CPU-based system, and GPU use is not strictly required.

\subsection{OTSU Binarisation}

The difference image is then binarised using the Otsu binarisation method (see \cite{otsu1979}). Otsu's method is particularly effective for images with bimodal histograms, where two distinct peaks represent two dominant pixel intensities, typically corresponding to the foreground and background. This method calculates a binarisation threshold which determines which pixels are black or white. White pixels, in this case, represent a large change of the pixel values which is caused by cloud movement, so a white pixel is seen as a cloud pixel whilst a black pixel represents "open sky" (no change). The moving telescope structure can either cause large differences in the pixel value and appears as cloud pixels (e.g. when slewing) or blocks the view to moving clouds having the structure appear as black pixels. As this is usually only visible in the pixels close to the horizon in areas where no observations are performed, these wrongly classified pixels do not pose a problem. As the position of the telescope is known at the time of the image analysis, the area around the structure in the cloud map can be masked and a flag can be saved alongside the masked pixels in the cloud map (see section \ref{sec:storagequery}).

The fundamental idea of the Otsu algorithm is to automatically find the intensity threshold that separates pixels into two classes (in our case, cloud and clear sky) by maximising the variance between these classes — that is, it selects the threshold where the difference between the average intensities of the two groups is greatest relative to their internal spread.

Otsu's method is applied to the difference image and thus separating pixels with high variability (cloud pixels) from pixels with low variability (open sky pixels). 

First, an image histogram with 256 bins is created using the graphics hardware. After this step, each bin is divided by the number of total pixels, in the image ($width\times height$) resulting in the normalised histogram (NH). The normalised histogram represents the probability of a pixel falling into a specific bin.

The next steps are then performed on the CPU for the 512-bin histogram, as these steps are not well-suited for parallel execution. Although it is possible to implement them on the GPU, the gain in speed is negligible compared to the overhead of compute shader setup and data transfer. 

Let the image histogram be binned into \( L \) discrete intensity levels (typically 256 or 512 bins). We denote by NH(i) the normalised histogram value at bin i, i.e. the fraction of all pixels in the image that have intensity i. For each candidate threshold bin \( k \), the histogram is split into two classes: \( C_0 \), containing intensity levels \( [1, \ldots, k] \), and \( C_1 \), containing levels \( [k+1, \ldots, L] \). The goal is to find the threshold bin \( k^* \) that maximises the between-class variance \( \sigma_B^2(k) \), providing the best separation between cloud and clear-sky pixels in the difference image.

Otsu then gives the formulae for the class probabilities:

\[
\omega_0(k) = \sum_{i=1}^k \operatorname{NH}(i)
\]
\[
\omega_1(k) = \sum_{i=k+1}^L \operatorname{NH}(i) = 1 - \omega_0(k)
\]

and the class mean intensities:

\[
\mu_0(k) = \frac{ \sum_{i=1}^k i \cdot \operatorname{NH}(i) }{ \omega_0(k) }
\]
\[
\mu_1(k) = \frac{ \sum_{i=k+1}^L i \cdot \operatorname{NH}(i) }{ \omega_1(k) }
\]

The between-class variance is then defined (see \cite{otsu1979}) as:

\[
\sigma_B^2(k) = \omega_0(k) \, \omega_1(k) \, \left( \mu_1(k) - \mu_0(k) \right)^2
\]

The optimal threshold \( k^* \) is the value of \( k \in [1, L-1] \) that maximises this variance:

\[
k^* = \max_{1 \leq k < L} \sigma_B^2(k)
\]

Once the optimal threshold \( k^* \) is determined, it is applied to the difference image to produce a binary cloud map. Pixels with values greater than or equal to \( k^* \) are classified as ``cloud'', and those below as ``clear sky''. 

In some cases, the lighting conditions between two adjacent images can change drastically, for example, when the telescope structure moves between the camera and the moon. 

To allow proper classification in these cases, an analysis was conducted over the time span from 2019 to 2023, analysing the calculated thresholds for different moon phases (lighting conditions). This resulted in a function which provides boundary conditions for the binarisation threshold of two adjacent cloud maps. The calculated threshold of a new cloud map is then clamped against these boundaries based on the previous cloud map's threshold to avoid misclassification of sky and cloud pixels in the difference images.

\subsection{Mathematical Morphology}
\label{subsec:morphology}
Stars with magnitudes below 3 mag (V band) and other bright objects like Jupiter or even the ISS can lead to artefacts in the difference image, visible as small spots covering about $5\times 5$ pixels on average. These artefacts can be removed or reduced by using techniques called Mathematical Morphology \citep{serra1982} particularly erosion and dilation. The technique applies a structuring element (in this case a $3\times 3$ pixel square, other elements such as circles or ellipses are also possible but computationally more expensive) to a binary image. 
The erosion is then defined mathematically as:
\[
I \ominus E=\left\{z \in Z \mid B_z \subseteq I\right\}
\]
where $I$ denotes the binary image, $E$ the structuring element ($3\times 3$ pixel square), $Z$ an integer grid and $B_z$ the translation of $B$ by the vector $z$. 

The structuring element is applied to each pixel. For the presented algorithm it has the form of a $3\times 3$ pixel square with
\[
E=\begin{pmatrix}
1 & 1 & 1 \\
1 & 1 & 1 \\
1 & 1 & 1 
\end{pmatrix}
\]
The structural element can be seen as a small subset of the original binarised image centered at a pixel $B_z$ of this binary image. Then, the minimum of all pixel values of the binarised image, where the structural element has a value of $1$ is calculated and written to the output image. This is repeated for each pixel in the binarised image. An example OpenGL pixel shader is given in listing \ref{lst:erode}.

\begin{lstlisting}[caption={pixel shader for erosion},label=lst:erode]
#version 330 core
in vec2 fUv; // coordinate of current pixel
out vec4 FragColor; // output, 32bit RGB 
uniform sampler2D Tex; // input binary image
uniform float width; // width of imput image
uniform float height; // height of input image

const int frad = 1; // results in 3x3 square 

void main(void)
{
	float w = 1.0 / width;
	float h = 1.0 / height;
	float erode = 1.0; 
	for (int j = -frad; j <= frad; j++)
	{
		for (int i = -frad; i <= frad; i++)
		{
			float fi = float(i);
			float fj = float(j);
			erode = min(erode, texture2D(Tex, fUv.xy + vec2(fi * w, fj * h)).r);
		}
	}
	FragColor = vec4(erode, erode, erode, 1.0);
}
\end{lstlisting}

Similarly to erosion, the morphological operation of dilation can be defined as $I \oplus E$. The difference is that instead of the minimum of the pixel values (resulting in erosion), the maximum is used (resulting in dilation) so that the line 21 in listing \ref{lst:erode}: \verb|erode = min(erode, ...);| becomes \verb|dilate = max(dilate, ...);| renaming the variable `\textit{erode}' to `\textit{dilate}' respectively.

\begin{figure}
\centering
  \includegraphics[width=\linewidth]{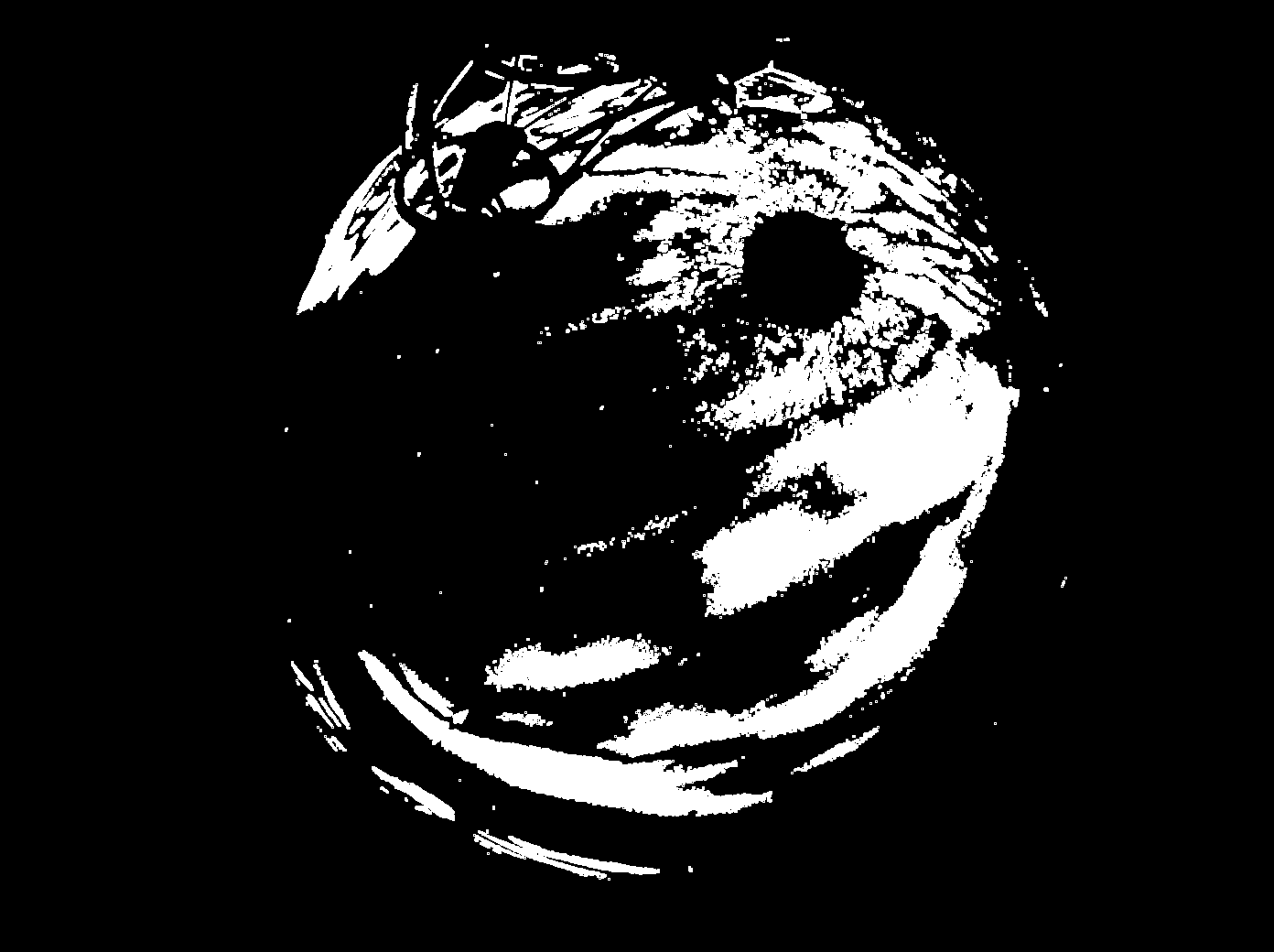}
  \caption{Binary image obtained after the application of the morphology methods (1) erosion and (2) dilation with a $3\times 3$ filter kernel. The number of star pixels is reduced whilst the cloud structure remains intact. Larger filter kernels (e.g. $5\times 5$ or $7\times 7$ will remove all stars but also small cloud structures like the ones present in the middle of the image}
  \label{fig:binaryfiltered}
\end{figure}

The result of this morphological process, applied to an unfiltered image can be seen in figure \ref{fig:binaryfiltered}. In this image, erosion is applied to the unfiltered image and thereafter, dilation is applied to the result of the erosion step. 

\subsection{Further Filtering}

The amount of visible bright stars in the image is reduced by using the technique in section \ref{subsec:morphology}, but some star-like structures still remain. The method can be improved for higher accuracy by applying a larger structural element (i.e. a $5\times 5$ or $7\times 7$ pixel square) but this will also affect the actual pictured cloud structure. To maintain the speed and keep the implementation simple and applicable to GPU hardware, a simple heuristical component detection algorithm can be applied after the erosion-dilation steps. 
The idea is to leverage the fast memory of modern graphics hardware and - for each pixel in the original image - analyse an area of $21\times 21$ pixels (with the current pixel being in the centre of this area) and perform the heuristical component detection as follows:
The algorithm analyses each row and then each column. If a row (or column) contains a switch from 0 to 1 (or 1 to 0) in pixel values, a marker is set stating that the row (or column) contains pixels belonging to one or more components. If a row (or column) exists between two marked rows (or columns) that only contains $0$-valued (dark) pixels, it is assumed that the area contains more than one component and is discarded. 
If only one component is found in the area \textbf{and} the component is fully included in the area (i.e. not at the edge) and the area of this component is below a threshold (20 pixels), the pixels of this area are set to 0. 
Although there might be cases of oddly shaped areas where the algorithm cannot find a single row or column (e.g. because the areas have to be split by a diagonal), it will reduce the amount of stars visible in the cloud pattern to almost zero and provide a simple and fast approximation to the ``optimal'' solution with all stars removed. This is because stars appear as small, convex shapes in elliptical form, so large groups of pixels with value 1 can be attributed to clouds.

\begin{figure}
\centering
  \includegraphics[width=\linewidth]{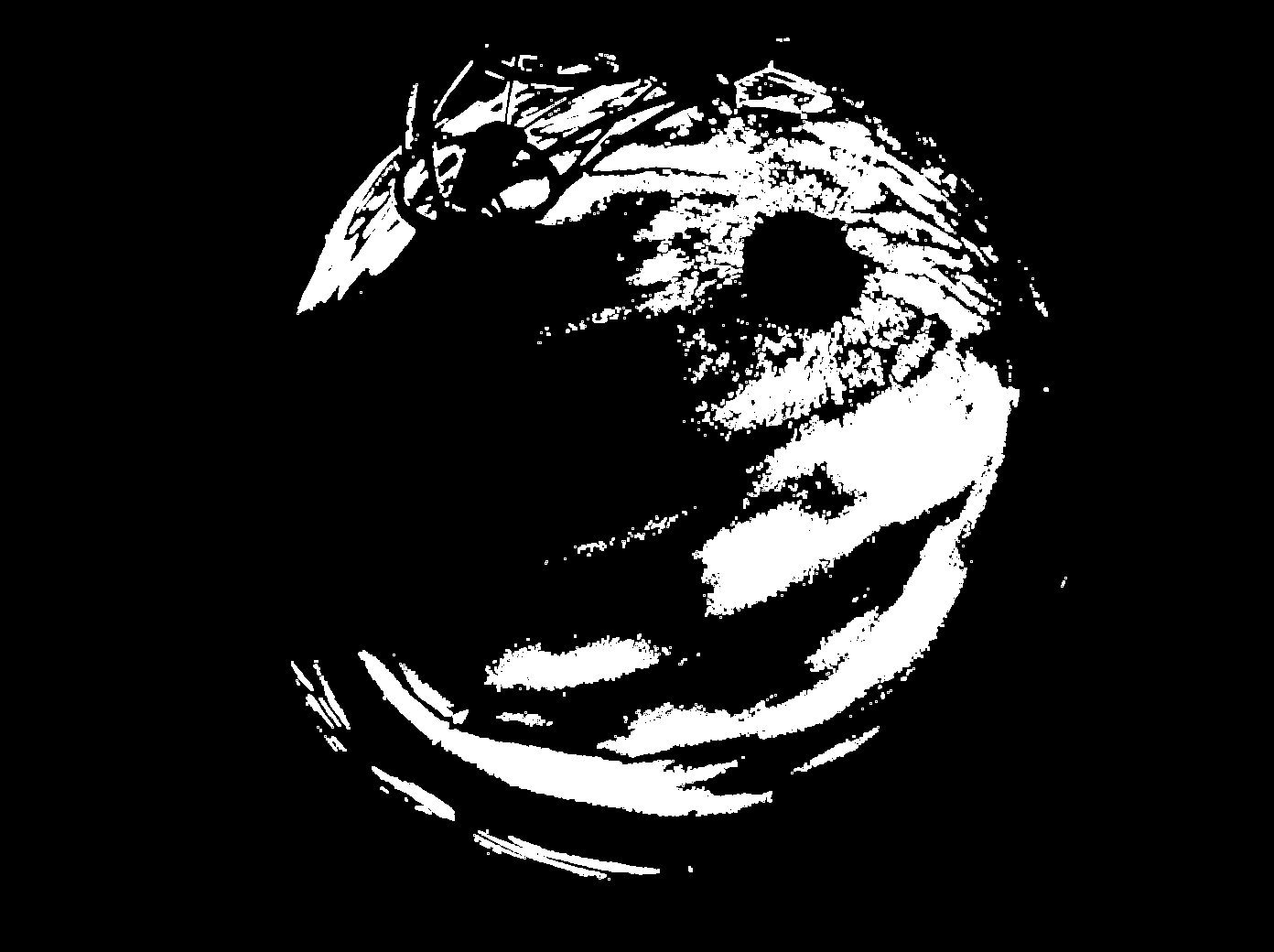}
  \caption{Binary image obtained after an additional star removal step, a simple heuristic is applied to a $21\times 21$ pixel window moved over the whole image, again leveraging fast GPU hardware, to test for small star-shaped pixel areas}
  \label{fig:posterodefilter}
\end{figure}

The result can be seen in figure \ref{fig:posterodefilter}.

\subsection{Storage and Query}
\label{sec:storagequery}
The generated cloud map is then stored as a FITS file with the world-coordinate system from the initial all-sky image on the file system. The difference image is stored in a similar way. Meta-information for this cloud-map is stored in a SQL Database. This includes the timestamp of the cloud image, the OTSU threshold and a parameter called the ``\textit{total cloudiness}'' which is simply the number of cloud pixels over the total number of pixels in the area of interest (in the case of the LT implementation, the area of interest is considered to be all pixels within a circle around the centre of the image corresponding to an altitude of $20^\circ$. 

\section{Cloud Tracking and Prediction}
\label{sec:cloudmovementprediction}
\subsection{Individual Cloud Identification}
While cloud detection is helpful to determine the current sky cover and check if an observation would be feasible \textit{now}, cloud tracking tries to identify individual clouds in the image and track their movement over time. This will allow to predict their location in the near future and support \textit{planning ahead} to help schedule observations which will be in areas of the sky where there will be no clouds within a certain time span. Any prediction will inevitably decrease in accuracy as the time span increases. Most Liverpool Telescope observation groups are relatively short ($\approx 15$min, including overheads) so this time has been adopted as a baseline for making predictions. This effectively can give the robotic scheduler the ability to decide whether a group is better observed \textit{now} or \textit{next} and the decision can be revisited after each observation is completed. 

The wind in the cloud layers of the atmosphere remains constant and does not suddenly change its direction, meaning clouds move in a constant direction for a relatively long period of time. Analysing the Liverpool Telescope Skycam A all-sky movies and conducting own experiments using an all-sky camera in Emleben (Thuringia, Germany), the shortest time for the general cloud movement direction changing by more than $60\deg$ was about two to three hours for wind speeds at which observations can be safely conducted. The actual speed at which the clouds are moving varies in similar time scales. It is thus acceptable to assume a constant cloud velocity vector for a prediction time of 20 to 30 minutes. 

In a first step, the fisheye projection of the all-sky camera has to be removed, the image has to be unprojected as the fisheye effect can be described as a projection of the full hemisphere over the camera into a circular region of the sensor. This un-projection is necessary as the clouds will not travel in straight paths through the original images but the cloud path will appear warped and bent. Also the shape of the cloud is distorted based on the position in the original image.

As the all-sky image covers a $180^\circ$ field of view, pixels near the horizon correspond to very large physical areas on the ground or in the lower atmosphere (up to hundreds of square meters per pixel), while pixels near the zenith cover much smaller physical areas. This effect arises from the projection geometry, as the angular resolution per pixel (in square degrees) is roughly uniform, but the physical footprint increases dramatically toward the horizon. The de-fishing process leads to over- and under-representation in the unprojected image. To account for this, cloud identification is limited to altitudes between $20^\circ$ and $90^\circ$.

To detect and identify clouds in the un-projected binary image, a connected-component labelling (CCL) algorithm using directed acyclic graphs, called Spaghetti Labelling as described in \citep{bolelli2020}, is used. \cite{bolelli2021} propose a GPU implementation for this algorithm but as the CPU implementation is already very efficient, the algorithm described in this paper does perform the CCL on the CPU. 

The CCL method identifies and labels the cloud areas which form single, connected areas in the image, as well as determining their centroid and their bounding box in a binary cloud pattern image. 

To speed up the following processes, identified cloud patches with an area below a set threshold (in this case 40 pixels) are discarded. In a future version it is planned to test if these small patches can be merged with nearby larger patches.

The information (label number, centroid coordinates, bounding box and area size) for each individual cloud patch is then stored. This process is repeated for every cloud map in a sequence.

\begin{figure}
\centering
  \includegraphics[width=\linewidth]{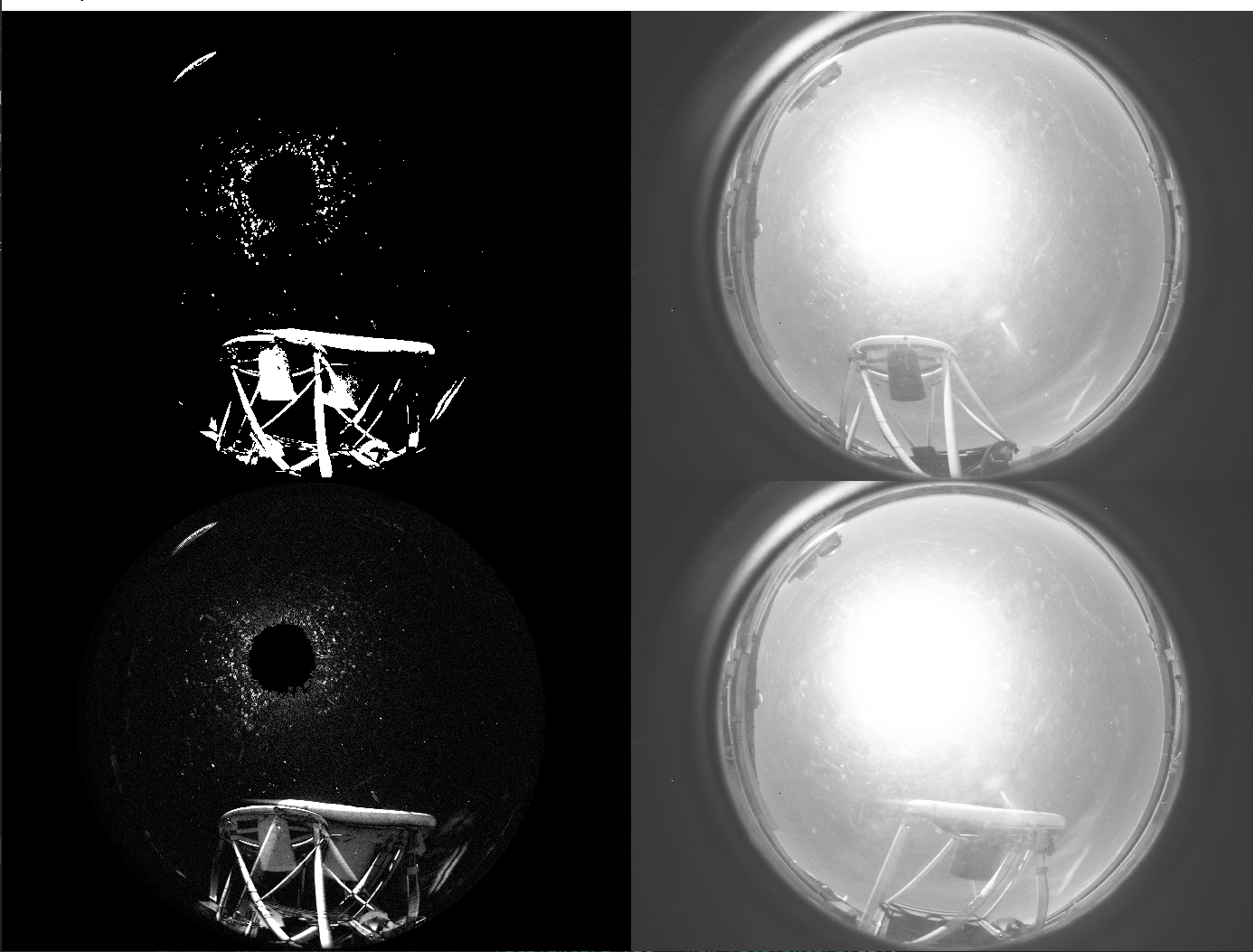}
  \caption{The fast slewing of the telescope and the full moon on the telescope structure caused the algorithm to detect the structure as a quite large cloud. The resulting cloud map can not be used for large parts of the sky. To detect this issue, large, sudden appearances of ``cloud-like structures'' will be discarded}
  \label{fig:slewing}
\end{figure}

One problem arising was that parts of the moving telescopes and the moon (if visible) were also detected as cloud-like structures and marked in the binary image due to the nature of the cloud detection algorithm (see section \ref{sec:clouddetection}). As this is usually not a problem for cloud lookups to detect clouds at a specific moment, it might actually lead to false results further in the cloud tracking, e.g. if a part of the moving telescope (see figure \ref{fig:slewing}) is labelled as cloud and moving to the opposite direction as the clouds. 
The position of the moon and its phase can be determined using well-known astronomical algorithms as described in \cite{meeus1998}. To keep the box of the telescope small, the current azimuth and altitude of the telescope must be known to the algorithm to shrink the affected area to a minimum. This is currently done using the world-coordinate system of images recorded by Skycam T, a piggy-back small telescope attached to the structure of the Liverpool Telescope (see \cite{mawson2013}).
Cloud centroids which are within the moon- or telescope area will be discarded as well. 

\begin{figure}
\centering
  \includegraphics[width=\linewidth]{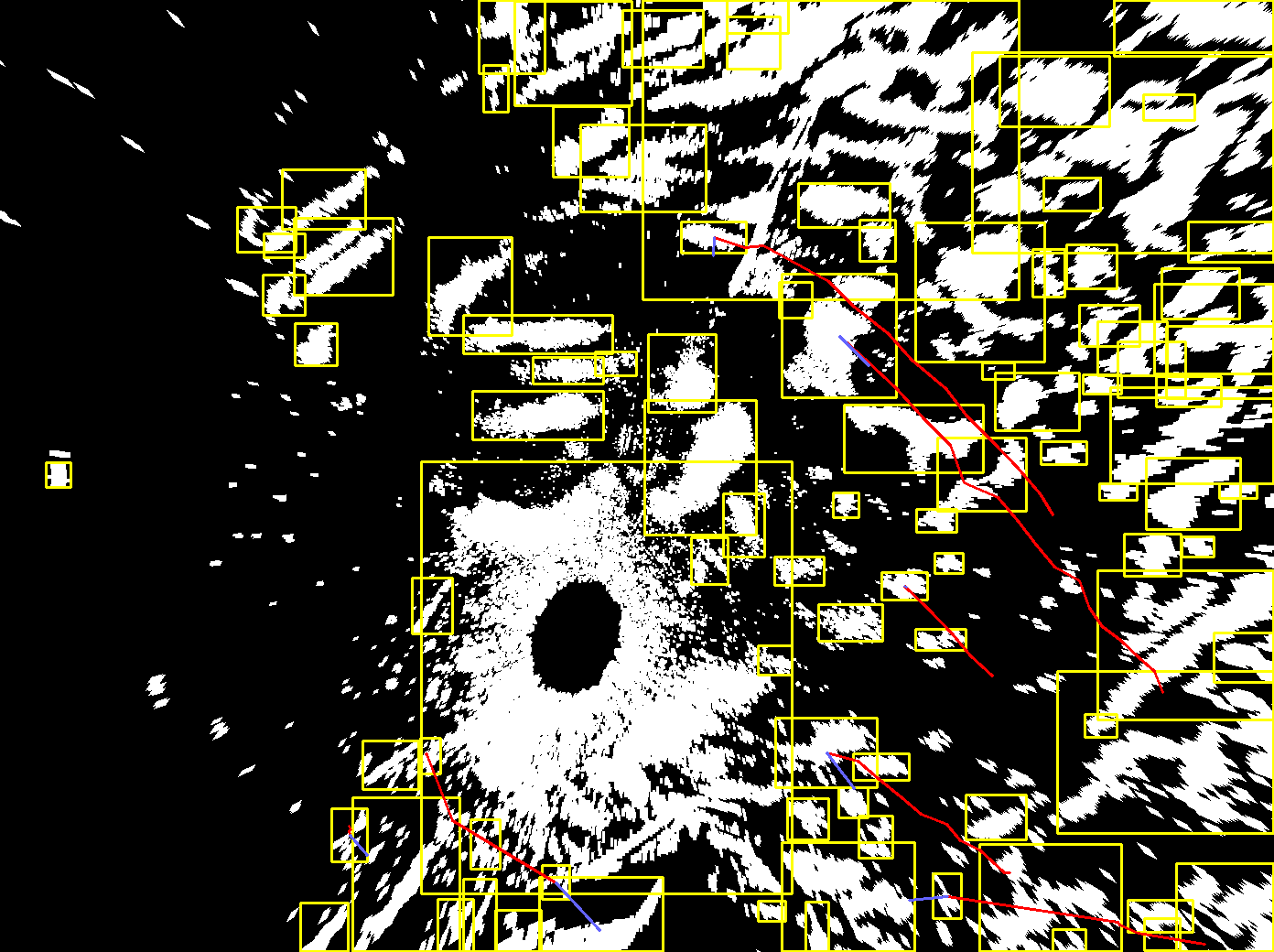}
  \caption{Detected cloud-like structures with their tracks in a cloud-map with the fish-eye effect removed. The yellow boxes mark the detected cloud-like structures in the image and the associated track for each detected cloud (the position of the centroid in the previous images) is displayed. It is also visible that the structure caused by dust and the over-exposure around the full Moon is vlearly visible and detected as a cloud-like structure. As the resulting path of this structure differs from the other cloud paths, it is discarded }
  \label{fig:alltracks}
\end{figure}

Figure \ref{fig:alltracks} shows a cloud pattern (dark red) with the detected cloud patches and their bounding boxes (yellow). 
\subsection{Cloud Tracking}

The detected cloud patches are then matched to clouds detected in the previous cloud pattern images using a metric $d$ based on the euclidean distance of the centroids and the overlapping area of the bounding box ($c_{x,y}$ representing the centroid position in pixel coordinates and $A_n$ the area of the cloud in pixels). 

\begin{equation}
d = \sqrt{(c_{x_2}-c_{x_1})^2+(c_{y_2}-c_{y_1})^2}+\frac{A_1+A_2}{Overlap(A_1,A_2)}
\end{equation}

This is done by evaluating the metric for each cloud patch in the current image against all cloud patches of the previous image. A `match' between two clouds is found (e.g. the two cloud patches represent the same cloud in two adjacent images) for the patches resulting in the smallest value of $d$. To avoid errors $d$ is also checked against a threshold. If $d$ is larger than this threshold, the match is discarded as the distance between the cloud patches is too large and thus cloud patches are considered independent. Tests of various thresholds have shown that a threshold of 200 is a good estimate for the cloud system at the Liverpool Telescope.

The coordinates of the centroids for each matching cloud component are then added to a data structure called 'track'. If no matching track is found, a new track is created. This data structure contains the coordinates of the centroids of all matched cloud patches in chronological order. 

The track itself is then formed of the path described by concatenating the vectors from centroid point $n$ to centroid point $n+1$, starting at point $1$. 

All tracks are then analysed and a metric for the ``smoothness'' of the track is determined. This smoothness is based on the angle between two adjacent vectors (which must not be larger than a threshold, here $10\deg$ is used). This can be efficiently calculated using the inner product. 

In this smoothing step, tracks which contain angles above the threshold are not considered ``cloud paths'' and ignored. The remaining tracks are then ``smoothed'' using a Gaussian filter (see \cite{shapiroComputerVision2001}).

\subsection{Cloud position prediction}
 Each track is then fed into a separate Kalman filter (see \cite{kalman1960}) to predict its future direction. A Kalman filter - also known as linear quadratic estimation - is an algorithm for estimating unknown variables using a series of measurements observed over time, including statistical noise and other inaccuracies.

The predicted direction vectors produced by the Kalman filter for each cloud track are averaged to estimate the overall cloud motion, resulting in a mean velocity vector $\vec{v}$. Typically, 5 to 10 consecutive cloud patterns are sufficient to make a reliable prediction. To ensure robustness, the averaging process includes a two-pass approach: first, the general direction of motion is determined, and second, the magnitude of $\vec{v}$ is refined by weighting vectors pointing within $\pm10^\circ$ of the estimated direction more heavily. The final displacement vector $\vec{v}$ is then applied to each cloud pixel in the unprojected all-sky cloud map to predict its location $n$ minutes into the future. This calculation can be efficiently performed on the GPU using a shader program.

A GPU pixel shader is then used to shift the detected cloud blobs by that vector $\vec{v}$ to determine the cloud map at time $t+1$. This is repeated for each minute in the future until the maximum prediction time (usually 20 minutes) is reached. 

A problem occurring is that the amount and location of clouds coming from the horizon is unknown. Due to the resolution of the camera and the angle, only a few pixels of the camera cover a large area of the sky (under-sampling) and thus reliable cloud information can not be retrieved. 

\section{Results and Discussions}
\label{sec:rd}

\subsection{Cloud Detection}

In order to test the efficiency of this cloud detection algorithm, it was tested against two samples in the historical dataset of Skycam-A-images. The first sample was obtained from February, 20\textsuperscript{th} 2019 to December 31\textsuperscript{st} 2022 and the second sample was obtained from January, 1\textsuperscript{st} 2024 to January 30\textsuperscript{th} 2024. The first sample consisted of 619,421 images and the second sample consisted of 16,937 images. Over these periods a variety of degrees of cloud cover were observed, from nights where conditions were fully photometric to nights where no stars were visible at all. The second sample was chosen as due to changes of our satellite image provider, in that time span the Liverpool Telescope operated without an external cloud sensor and thus was operating in cloud conditions that normally would have been resulted in observations being suspended. 

The algorithm was tested against \citep{adam2017} (see section \ref{sec:intro}) who generously provided their code on Github (\url{https://github.com/tudo-astroparticlephysics/starry_night}) as well as a method using the Skycam T at the Liverpool Telescope.

The comparison of \citep{adam2017} against the algorithm presented in this paper, both using the Skycam A, could only be performed in dark nights when there was no visible moon. During times with visible moon, especially with a moon disk of $30\%$ or more, their algorithm was unable to perform photometry in large areas of the sky. With the full moon up, the method did not provide any useful results at all. This is caused by hardware limitations of the all-sky camera equipment at the Liverpool Telescope (Starlight Xpress Oculus). However, the problem of over-exposed areas around the visible moon is a common problem on most typical all-sky cameras 

In dark nights both algorithms matched against each other by around $92\%$ of detected cloud area, depending on the density of catalogue stars in the area of the sky. A limitation of their algorithm is the fact that it can only outline clouds based on visible and non-visible stars in a provided catalogue whilst the method presented in this paper is able to accurately detect the shape of the cloud which is the main cause for the average match of `only' $92\%$. 

\begin{figure}
\centering
  \includegraphics[width=\linewidth]{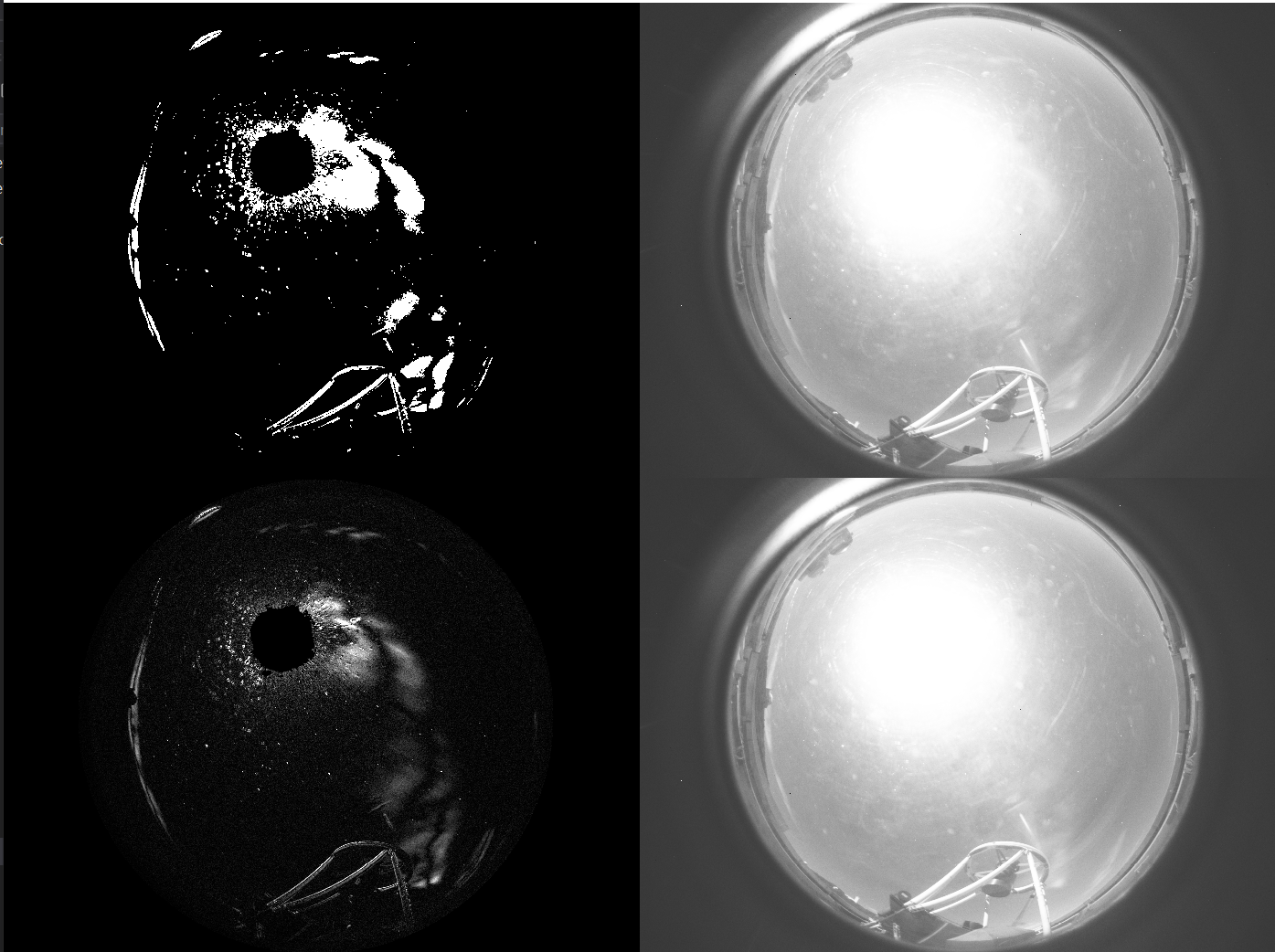}
  \caption{Principle work of the algorithm: The right column shows the input images at $t$ and $t+1min$, the resulting difference image is shown on the bottom left and finally, the binarised cloud map is shown in the top left}
  \label{fig:nearmoon}
\end{figure}

In bright nights (moon disk $>30\%$), whilst their algorithm failed to provide a cloud map (using the images recorded with the equipment at the Liverpool Telescope), the method presented in this paper was able to accurately detect clouds around the actual moon disk. In figure \ref{fig:nearmoon}, the right side represents two all-sky images taken one minute apart. The bottom left side is the difference image and the top left picture contains the extracted cloud pattern. The cloud pattern contains various small white patches falsely identified as clouds. These are the result of moonlight scattering on dust accumulated on the camera protection over time. With an exposure time of $30$s, the moon disk itself saturates the CCD completely and the ADU values in the pixel regions around the moon disk are also close to saturation levels resulting in visible noise. This noise is falsely labelled as a cloud but it has little to no influence on the observations itself as observations this close to the moon are very rare. 

The next test was conducted with the help of our Skycam T. Skycam T is a Starlight Xpress Trius SX-35 camera mounted on the top-end of the Liverpool Telescope and parallel points with the telescope. The camera currently uses a Zeiss Planar T 85mm f/1.4 ZF2 lens which results in a pixel scale of about 44 arcsec and a field-of-view of roughly $24\times16^\circ$ and a detection limit of about 12mag. Skycam T takes a 10 second exposure every minute and stores it in a public available database on the website of the Liverpool Telescope. All Skycam T images are plate-solved using the Astromentry.net software suite \citep{lang2010} and a world coordinate system (WCS) is added to the FITS file. 

To use this system to test the cloud detection system, a photometric analysis of the Skycam T images using the SExtractor tool \citep{bertin1996} was performed. The images of cloud-free nights were analysed with the APASS catalogue \citep{henden2018}. APASS stars are extracted from the images to calibrate the instrument magnitude. 

In the next step, nights with clouds were randomly selected from the database. Then, for every Skycam T image in these nights, the sky location of the centre pixel (in terms of Right Ascension and Declination) was determined and projected into the cloud map image and an area of about $15^\circ$ around this pixel is then read from the cloud map and the ``cloudiness'' parameter is calculated as described in section \ref{sec:storagequery}. 

For the all-sky camera, we determine which pixels correspond to the Skycam-T field of view. We then calculate the `cloudiness' parameter, which is the fraction of these pixels which are classified as cloud using our detection algorithm. For Skycam T, the APASS database is queried for the same section of the sky, and all stars with magnitudes smaller than 10mag are retrieved. The number of stars in the database query result determines the stars that \textit{should} be visible (expected number of stars), which is then compared to the photometric analysis performed with SExtractor, determining the stars that actually \textit{are} visible. We consider a region in the all-sky cloud map to be cloud-covered if more than 75\% of the pixels in that region are classified as cloud. This threshold mirrors the classification approach used for Skycam T, allowing for direct comparison between both systems.

If the number of visible stars is below $75\%$ of the expected number of stars, the image is labelled as ``cloud covered''. The threshold of $75\%$ was chosen due to the optical limitations of Skycam T. As it is a medium-field camera with a relatively large field-of-view and SExtractor only performs aperture photometry, especially in areas with a high star density (crowded fields) SExtractor might not be able to extract all APASS stars without using more sophisticated techniques like PSF photometry. To account for thin clouds and to calculate the detection threshold of the cloud system, a second analysis is performed if the detected number of starts is above the $75\%$ threshold of the expected stars. Here, the measured star magnitudes are compared against the catalogue values (using the calibration performed on a clear night) and if the average magnitude is below 3mag of the catalogue value, the image is also labelled as ``cloud covered''. Clouds which dim the bright APASS stars with less than 10mag by at least 3mag are also detected by the all-sky cloud monitor.

For the cloud detection evaluation, 47 nights from 2019 to 2023 were selected to represent a wide range of observing conditions, including bright nights with and without the Moon, as well as dark nights. These nights were identified via entries in the night log\footnote{\url{https://telescope.ljmu.ac.uk/Reports/}} and visual inspection of the nightly all-sky video sequences\footnote{\url{https://telescope.livjm.ac.uk/data/webfiles/Skycam/browse2025.html}}. From a total of 32,307 all-sky images across these nights, the first and last 60 minutes of each night were excluded to remove twilight effects. From the remaining images, one 100-image set and two 1000-image sets were randomly selected:

\begin{itemize}
    \item \textbf{Group 1}: 100 images from bright nights (Moon altitude $> -5^\circ$, Moon disk $>10\%$) where the angular separation between the target and the Moon was less than $25^\circ$. This configuration is relatively rare.
    \item \textbf{Group 2}: 1000 images from bright nights with Moon–target angular separation greater than $25^\circ$.
    \item \textbf{Group 3}: 1000 images from dark nights (Moon altitude $< -5^\circ$ and Moon disk $<10\%$).
\end{itemize}

Testing the cloud detection system using the Skycam T method, the following results were obtained:

For bright nights we have a false negative rate (the sky area in the cloud map is labelled ``clear'' whilst is is actually cloud covered) of less than $1\%$ of the analysed images and a false positive rate also of $1\%$ outside a radius of $25^\circ$ of the moon disk. Inside this area, the false positive rate can climb up to $10-15\%$ (especially with a visible moon disk of more than $50\%$) due to moonlight scattering on dirt and dust on the sensor cover as well as lens-flare effects. As astronomical observations are rarely performed so close to the moon, this relatively high false positive rate is of no concern. The rate could be improved by frequently cleaning the camera's protective elements and using a sensor with a higher dynamic range. 
\setlength{\tabcolsep}{2pt} 
\begin{table}
\centering
\begin{tabular}{ccc}

\rowcolor{Bisque!50}\textbf{Bright $>25^\circ$} & \textbf{Bright $<25^\circ$} & \textbf{Dark Nights} \\
\begin{tabular}{lcc}
 & C & O \\
C & 0.46 & 0.01 \\
O & 0.01 & 0.52 \\
\rowcolor{Bisque!50}A & & 0.98 \\
\rowcolor{Bisque!50}P & & 0.97
\end{tabular}
&
\begin{tabular}{lcc}
 & C & O \\
C & 0.41 & 0.13 \\
O & 0.00 & 0.46 \\
\rowcolor{Bisque!50}A & & 0.87 \\
\rowcolor{Bisque!50}P & & 0.76
\end{tabular}
&
\begin{tabular}{lcc}
 & C & O \\
C & 0.41 & 0.01 \\
O & 0.14 & 0.44 \\
\rowcolor{Bisque!50}A & & 0.85 \\
\rowcolor{Bisque!50}P & & 0.97
\end{tabular}
\\

\end{tabular}
\setlength{\tabcolsep}{6pt} 
\caption{
Confusion matrices for cloud detection under three observing conditions. Each matrix is normalized such that values sum to 1.0. Rows correspond to \textbf{predicted} class, columns to \textbf{actual} class. C = \textbf{C}loud, O = \textbf{O}pen sky. “Bright $<25^\circ$” refers to bright nights with the target close to the Moon (rare); “Bright $>25^\circ$” to nights with large Moon-target separation; and “Dark Nights” to low-illumination cases. Group sizes: 100 (Bright $<25^\circ$), 1000 (others). The bottom rows of the table display \textbf{A}ccuracy and \textbf{P}recission. All values rounded to two decimals. 
}

\label{tab:confmatrix_three}
\end{table}

In dark nights, the false positive rate is similar to the bright nights of about $1\%$. Relatively thin, large and slow-moving clouds can impose a problem on the false negative rate. Due to technical limitations of the camera, the contrast between the cloud and the open sky is too low to impact the difference image (see section \ref{sec:clouddetection}). In these cases, a false negative rate up to $15\%$ of the images is possible and was observed of a total of 17 nights in the years 2020 and 2021.

\subsection{Cloud Tracking}

\begin{figure*}
    \centering
    \begin{minipage}[b]{0.48\linewidth}
        \centering
            \begin{tikzpicture}
    \begin{groupplot}[
    group style={
        group name=my plots,
        group size=1 by 2,
        xlabels at=edge bottom,
        xticklabels at=edge bottom,
        vertical sep=5pt
    },
        width=\linewidth, 
        height=5cm,
        xlabel={Time (minutes)},
        ylabel={Cloudiness},
        xmin=0, xmax=30,
        ymin=0, ymax=1,
        grid=major,
        xtick={0, 5, 10, 15, 20, 25, 30},
        after end axis/.append code={\pgfextra{\vspace{-\baselineskip}}}
    ]
    \nextgroupplot
    \addplot[color=red, mark=*, mark options={fill=red, scale=0.6}] 
        coordinates {
 (0, 0.998) (1, 0.468) (2, 0.694) (3, 0.999) (4, 0.194) (5, 0) (6, 0.32) (7, 0.982) (8, 0.505) (9, 0.794) (10, 0.756) (11, 0.621) (12, 0.417) (13, 0.182) (14, 0.49) (15, 0.159) (16, 0.358) (17, 0.344) (18, 0.396) (19, 0.558) (20, 0.033) (21, 0.015) (22, 0.007) (23, 0.015) (24, 0.011) (25, 0.173) (26, 0.15) (27, 0.411) (28, 0.635) (29, 0.009) 
    };
    
    \addplot[color=blue, mark=square*, mark options={fill=blue, scale=0.6}] 
        coordinates {
  (0, 0.998) (1, 0.328) (2, 0.907) (3, 0.928) (4, 0.682) (5, 0.027) (6, 0.024) (7, 1) (8, 0.767) (9, 0.996) (10, 0.604) (11, 0.413) (12, 0.683) (13, 0.592) (14, 0.055) (15, 0.468) (16, 0.723) (17, 0.265) (18, 0.212) (19, 0.324) (20, 0.597) (21, 0.281) (22, 0.027) (23, 0.04) (24, 0.028) (25, 0.078) (26, 0.066) (27, 0.031) (28, 0.754) (29, 0.284) 
    };
    
    \addplot[color=green, mark=triangle*, mark options={fill=green, scale=0.6}] 
        coordinates {
(0, 0.998) (1, 0.998) (2, 0.998) (3, 0.998) (4, 0.998) (5, 0.998) (6, 0.998) (7, 0.998) (8, 0.998) (9, 0.998) (10, 0.998) (11, 0.998) (12, 0.998) (13, 0.998) (14, 0.998) (15, 0.998) (16, 0.998) (17, 0.998) (18, 0.998) (19, 0.998) (20, 0.998) (21, 0.998) (22, 0.998) (23, 0.998) (24, 0.998) (25, 0.998) (26, 0.998) (27, 0.998) (28, 0.998) (29, 0.998) 
    };
    \nextgroupplot[ymin=-1, ymax=1, height=3cm, ylabel={Residuals}, ytick={-1, -0.5, 0, 0.5, 1}, grid=major, xmajorgrids=false]
       \addplot[ycomb, color=orange, mark=*, mark options={fill=orange, scale=0.6}] 
        coordinates {
        (0, 0) (1, 0.14) (2, -0.213) (3, 0.071) (4, -0.488) (5, -0.027) (6, 0.296) (7, -0.018) (8, -0.262) (9, -0.202) (10, 0.153) (11, 0.208) (12, -0.266) (13, -0.41) (14, 0.435) (15, -0.309) (16, -0.365) (17, 0.079) (18, 0.184) (19, 0.234) (20, -0.564) (21, -0.267) (22, -0.019) (23, -0.025) (24, -0.017) (25, 0.095) (26, 0.084) (27, 0.38) (28, -0.119) (29, -0.274) 
    };
    
    \addplot[ycomb,color=purple, mark=square*, mark options={fill=purple, scale=0.6}] 
        coordinates {
(0, 0) (1, 0.67) (2, 0.091) (3, 0.07) (4, 0.316) (5, 0.972) (6, 0.975) (7, -0.001) (8, 0.232) (9, 0.002) (10, 0.394) (11, 0.586) (12, 0.316) (13, 0.407) (14, 0.943) (15, 0.53) (16, 0.275) (17, 0.734) (18, 0.786) (19, 0.674) (20, 0.401) (21, 0.717) (22, 0.972) (23, 0.959) (24, 0.97) (25, 0.92) (26, 0.932) (27, 0.967) (28, 0.244) (29, 0.715) 
    };
    \end{groupplot}
    \end{tikzpicture}

    \end{minipage}
    \hfill
    \begin{minipage}[b]{0.48\linewidth}
        \centering
            \begin{tikzpicture}
    \begin{groupplot}[
    group style={
        group name=my plots,
        group size=1 by 2,
        xlabels at=edge bottom,
        xticklabels at=edge bottom,
        vertical sep=5pt
    },
        width=\linewidth, 
        height=5cm,
        xlabel={Time (minutes)},
        ylabel={Cloudiness},
        xmin=0, xmax=30,
        ymin=0, ymax=1,
        grid=major,
        xtick={0, 5, 10, 15, 20, 25, 30},
        after end axis/.append code={\pgfextra{\vspace{-\baselineskip}}}
    ]
    \nextgroupplot
    \addplot[color=red, mark=*, mark options={fill=red, scale=0.6}] 
        coordinates {
(0, 0.696) (1, 0.775) (2, 0.849) (3, 0.902) (4, 0.816) (5, 0.182) (6, 0.309) (7, 0.454) (8, 0) (9, 0.022) (10, 0.964) (11, 0.976) (12, 0.016) (13, 0.025) (14, 0.965) (15, 0.061) (16, 0.054) (17, 0.218) (18, 0.751) (19, 0.61) (20, 0.593) (21, 0.439) (22, 0.443) (23, 0.344) (24, 0.352) (25, 0.297) (26, 0.288) (27, 0.255) (28, 0.238) (29, 0.244)     };
    
    \addplot[color=blue, mark=square*, mark options={fill=blue, scale=0.6}] 
        coordinates {
(0, 0.696) (1, 0.794) (2, 0.818) (3, 0.949) (4, 0.608) (5, 0.814) (6, 0.24) (7, 0.054) (8, 0) (9, 0.003) (10, 0.206) (11, 0.444) (12, 0.003) (13, 0.003) (14, 0.488) (15, 0) (16, 0.003) (17, 0.003) (18, 0.511) (19, 0.723) (20, 0.628) (21, 0.809) (22, 0.667) (23, 0.694) (24, 0.525) (25, 0.696) (26, 0.703) (27, 0.582) (28, 0.659) (29, 0.705) 
    };
    
    \addplot[color=green, mark=triangle*, mark options={fill=green, scale=0.6}] 
        coordinates {
(0, 0.696) (1, 0.696) (2, 0.696) (3, 0.696) (4, 0.696) (5, 0.696) (6, 0.696) (7, 0.696) (8, 0.696) (9, 0.696) (10, 0.696) (11, 0.696) (12, 0.696) (13, 0.696) (14, 0.696) (15, 0.696) (16, 0.696) (17, 0.696) (18, 0.696) (19, 0.696) (20, 0.696) (21, 0.696) (22, 0.696) (23, 0.696) (24, 0.696) (25, 0.696) (26, 0.696) (27, 0.696) (28, 0.696) (29, 0.696) 
    };
    \nextgroupplot[ymin=-1, ymax=1, height=3cm, ylabel={Residuals}, ytick={-1, -0.5, 0, 0.5, 1}, grid=major, xmajorgrids=false]
       \addplot[ycomb, color=orange, mark=*, mark options={fill=orange, scale=0.6}] 
        coordinates {
(0, 0) (1, -0.019) (2, 0.031) (3, -0.046) (4, 0.208) (5, -0.632) (6, 0.069) (7, 0.4) (8, 0) (9, 0.018) (10, 0.757) (11, 0.532) (12, 0.013) (13, 0.022) (14, 0.477) (15, 0.061) (16, 0.05) (17, 0.215) (18, 0.239) (19, -0.113) (20, -0.035) (21, -0.37) (22, -0.225) (23, -0.351) (24, -0.173) (25, -0.399) (26, -0.415) (27, -0.327) (28, -0.421) (29, -0.461) 
    };
    
    \addplot[ycomb,color=purple, mark=square*, mark options={fill=purple, scale=0.6}] 
        coordinates {
(0, 0) (1, -0.098) (2, -0.122) (3, -0.252) (4, 0.089) (5, -0.117) (6, 0.457) (7, 0.642) (8, 0.696) (9, 0.693) (10, 0.49) (11, 0.252) (12, 0.693) (13, 0.693) (14, 0.208) (15, 0.696) (16, 0.693) (17, 0.693) (18, 0.185) (19, -0.027) (20, 0.068) (21, -0.113) (22, 0.029) (23, 0.002) (24, 0.171) (25, 0) (26, -0.007) (27, 0.114) (28, 0.037) (29, -0.009) 
    };
    \end{groupplot}
    \end{tikzpicture}

    \end{minipage}
    \begin{minipage}[b]{0.48\linewidth}
        \centering
            \begin{tikzpicture}
    \begin{groupplot}[
    group style={
        group name=my plots,
        group size=1 by 2,
        xlabels at=edge bottom,
        xticklabels at=edge bottom,
        vertical sep=5pt
    },
        width=\linewidth, 
        height=5cm,
        xlabel={Time (minutes)},
        ylabel={Cloudiness},
        xmin=0, xmax=30,
        ymin=0, ymax=1,
        grid=major,
        xtick={0, 5, 10, 15, 20, 25, 30},
        after end axis/.append code={\pgfextra{\vspace{-\baselineskip}}}
    ]
    \nextgroupplot
    \addplot[color=red, mark=*, mark options={fill=red, scale=0.6}] 
        coordinates {
(0, 0.83) (1, 0.797) (2, 0.835) (3, 0.773) (4, 0.724) (5, 0.483) (6, 0.428) (7, 0.244) (8, 0.576) (9, 0.476) (10, 0.509) (11, 0.113)
    };
    
    \addplot[color=blue, mark=square*, mark options={fill=blue, scale=0.6}] 
        coordinates {
  (0, 0.83) (1, 0.781) (2, 0.788) (3, 0.804) (4, 0.774) (5, 0.566) (6, 0.536) (7, 0.342) (8, 0.782) (9, 0.832) (10, 0.808) (11, 0.341) (12, 0.495) (13, 0.816) (14, 0.724) (15, 0.781) (16, 0.782) (17, 0.783) (18, 0.724) (19, 0.643) (20, 0.617) (21, 0.545) (22, 0.496) (23, 0.508) (24, 0.635) (25, 0.652) (26, 0.35) (27, 0.058) (28, 0.053) (29, 0.627) 
    };
    
    \addplot[color=green, mark=triangle*, mark options={fill=green, scale=0.6}] 
        coordinates {
 (0, 0.83) (1, 0.83) (2, 0.83) (3, 0.83) (4, 0.83) (5, 0.83) (6, 0.83) (7, 0.83) (8, 0.83) (9, 0.83) (10, 0.83) (11, 0.83) (12, 0.83) (13, 0.83) (14, 0.83) (15, 0.83) (16, 0.83) (17, 0.83) (18, 0.83) (19, 0.83) (20, 0.83) (21, 0.83) (22, 0.83) (23, 0.83) (24, 0.83) (25, 0.83) (26, 0.83) (27, 0.83) (28, 0.83) (29, 0.83) 
    };
    \nextgroupplot[ymin=-1, ymax=1, height=3cm, ylabel={Residuals}, ytick={-1, -0.5, 0, 0.5, 1}, grid=major, xmajorgrids=false]
       \addplot[ycomb, color=orange, mark=*, mark options={fill=orange, scale=0.6}] 
        coordinates {
        
(0, 0) (1, 0.016) (2, 0.047) (3, -0.032) (4, -0.049) (5, -0.083) (6, -0.108) (7, -0.098) (8, -0.206) (9, -0.356) (10, -0.299) (11, -0.228)     };
    
    \addplot[ycomb,color=purple, mark=square*, mark options={fill=purple, scale=0.6}] 
        coordinates {
        (0, 0) (1, 0.05) (2, 0.042) (3, 0.026) (4, 0.057) (5, 0.265) (6, 0.295) (7, 0.489) (8, 0.048) (9, -0.001) (10, 0.022) (11, 0.49) (12, 0.335) (13, 0.015) (14, 0.106) (15, 0.049) (16, 0.049) (17, 0.048) (18, 0.106) (19, 0.187) (20, 0.213) (21, 0.286) (22, 0.335) (23, 0.322) (24, 0.195) (25, 0.179) (26, 0.48) (27, 0.772) (28, 0.777) (29, 0.203) 
    };
    \end{groupplot}
    \end{tikzpicture}

    \end{minipage}
    \hfill
    \begin{minipage}[b]{0.48\linewidth}
        \centering
            \begin{tikzpicture}
    \begin{groupplot}[
    group style={
        group name=my plots,
        group size=1 by 2,
        xlabels at=edge bottom,
        xticklabels at=edge bottom,
        vertical sep=5pt
    },
        width=\linewidth, 
        height=5cm,
        xlabel={Time (minutes)},
        ylabel={Cloudiness},
        xmin=0, xmax=30,
        ymin=0, ymax=1,
        grid=major,
        xtick={0, 5, 10, 15, 20, 25, 30},
        after end axis/.append code={\pgfextra{\vspace{-\baselineskip}}}
    ]
    \nextgroupplot
    \addplot[color=red, mark=*, mark options={fill=red, scale=0.6}] 
        coordinates {
(0, 0.827) (1, 0.385) (2, 0.182) (3, 0.526) (4, 0.688) (5, 0.438) (6, 0.449) (7, 0.407) (8, 0.55) (9, 0.047) (10, 0.066) (11, 0.461) (12, 0.81) (13, 0.498) (14, 0.746)      };
    
    \addplot[color=blue, mark=square*, mark options={fill=blue, scale=0.6}] 
        coordinates {
(0, 0.827) (1, 0.609) (2, 0.404) (3, 0.141) (4, 0.302) (5, 0.291) (6, 0.141) (7, 0.204) (8, 0.215) (9, 0.151) (10, 0.043) (11, 0.179) (12, 0.473) (13, 0.83) (14, 0.867) (15, 0.672) (16, 0.095) (17, 0) (18, 0) (19, 0) (20, 0) (21, 0) (22, 0) (23, 0) (24, 0.004) (25, 0.013) (26, 0) (27, 0) (28, 0.015) (29, 0.007)
    };
    
    \addplot[color=green, mark=triangle*, mark options={fill=green, scale=0.6}] 
        coordinates {
(0, 0.827) (1, 0.827) (2, 0.827) (3, 0.827) (4, 0.827) (5, 0.827) (6, 0.827) (7, 0.827) (8, 0.827) (9, 0.827) (10, 0.827) (11, 0.827) (12, 0.827) (13, 0.827) (14, 0.827) (15, 0.827) (16, 0.827) (17, 0.827) (18, 0.827) (19, 0.827) (20, 0.827) (21, 0.827) (22, 0.827) (23, 0.827) (24, 0.827) (25, 0.827) (26, 0.827) (27, 0.827) (28, 0.827) (29, 0.827) 
    };
    \nextgroupplot[ymin=-1, ymax=1, height=3cm, ylabel={Residuals}, ytick={-1, -0.5, 0, 0.5, 1}, grid=major, xmajorgrids=false]
       \addplot[ycomb, color=orange, mark=*, mark options={fill=orange, scale=0.6}] 
        coordinates {
(0, 0) (1, -0.224) (2, -0.221) (3, 0.384) (4, 0.386) (5, 0.148) (6, 0.309) (7, 0.203) (8, 0.335) (9, -0.103) (10, 0.023) (11, 0.282) (12, 0.337) (13, -0.332) (14, -0.122) 
    };
    
    \addplot[ycomb,color=purple, mark=square*, mark options={fill=purple, scale=0.6}] 
        coordinates {
 (0, 0) (1, 0.219) (2, 0.424) (3, 0.686) (4, 0.526) (5, 0.537) (6, 0.687) (7, 0.624) (8, 0.613) (9, 0.677) (10, 0.785) (11, 0.649) (12, 0.355) (13, -0.002) (14, -0.04) (15, 0.155) (16, 0.732) (17, 0.827) (18, 0.827) (19, 0.827) (20, 0.827) (21, 0.827) (22, 0.827) (23, 0.827) (24, 0.823) (25, 0.814) (26, 0.827) (27, 0.827) (28, 0.813) (29, 0.82) 
    };
    \end{groupplot}
    \end{tikzpicture}

    \end{minipage}
    \begin{minipage}[b]{0.48\linewidth}
        \centering
            \begin{tikzpicture}
    \begin{groupplot}[
    group style={
        group name=my plots,
        group size=1 by 2,
        xlabels at=edge bottom,
        xticklabels at=edge bottom,
        vertical sep=5pt
    },
        width=\linewidth, 
        height=5cm,
        xlabel={Time (minutes)},
        ylabel={Cloudiness},
        xmin=0, xmax=30,
        ymin=0, ymax=1,
        grid=major,
        xtick={0, 5, 10, 15, 20, 25, 30},
        after end axis/.append code={\pgfextra{\vspace{-\baselineskip}}}
    ]
    \nextgroupplot
    \addplot[color=red, mark=*, mark options={fill=red, scale=0.6}] 
        coordinates {
(0, 0.007) (1, 0.003) (2, 0) (3, 0.003) (4, 0) (5, 0) (6, 0) (7, 0) (8, 0) (9, 0) (10, 0) (11, 0.006) (12, 0.595) (13, 0.936) (14, 1) (15, 0.264) (16, 0.005) (17, 0.007) (18, 0.01) (19, 0) (20, 0.003) (21, 0) (22, 0) (23, 0) (24, 0) 
    };
    
    \addplot[color=blue, mark=square*, mark options={fill=blue, scale=0.6}] 
        coordinates {
  (0, 0.007) (1, 0.006) (2, 0.002) (3, 0.006) (4, 0.002) (5, 0) (6, 0.002) (7, 0.004) (8, 0.002) (9, 0) (10, 0) (11, 0.001) (12, 0) (13, 0.722) (14, 0.884) (15, 1) (16, 0.513) (17, 0) (18, 0.001) (19, 0.002) (20, 0.004) (21, 0.002) (22, 0.002) (23, 0) (24, 0) (25, 0) (26, 0) (27, 0.067) (28, 0.441) (29, 0.645) 
    };
    
    \addplot[color=green, mark=triangle*, mark options={fill=green, scale=0.6}] 
        coordinates {
(0, 0.007) (1, 0.007) (2, 0.007) (3, 0.007) (4, 0.007) (5, 0.007) (6, 0.007) (7, 0.007) (8, 0.007) (9, 0.007) (10, 0.007) (11, 0.007) (12, 0.007) (13, 0.007) (14, 0.007) (15, 0.007) (16, 0.007) (17, 0.007) (18, 0.007) (19, 0.007) (20, 0.007) (21, 0.007) (22, 0.007) (23, 0.007) (24, 0.007) (25, 0.007) (26, 0.007) (27, 0.007) (28, 0.007) (29, 0.007) 
    };
    \nextgroupplot[ymin=-1, ymax=1, height=3cm, ylabel={Residuals}, ytick={-1, -0.5, 0, 0.5, 1}, grid=major, xmajorgrids=false]
       \addplot[ycomb, color=orange, mark=*, mark options={fill=orange, scale=0.6}] 
        coordinates {
(0, 0) (1, -0.003) (2, -0.002) (3, -0.002) (4, -0.002) (5, 0) (6, -0.002) (7, -0.004) (8, -0.002) (9, 0) (10, 0) (11, 0.005) (12, 0.595) (13, 0.214) (14, 0.116) (15, -0.736) (16, -0.508) (17, 0.007) (18, 0.009) (19, -0.002) (20, -0) (21, -0.002) (22, -0.002) (23, -0) (24, 0)
    };
    
    \addplot[ycomb,color=purple, mark=square*, mark options={fill=purple, scale=0.6}] 
        coordinates {
        (0, 0) (1, 0.001) (2, 0.004) (3, 0.001) (4, 0.005) (5, 0.007) (6, 0.004) (7, 0.003) (8, 0.004) (9, 0.007) (10, 0.007) (11, 0.005) (12, 0.007) (13, -0.715) (14, -0.877) (15, -0.993) (16, -0.506) (17, 0.007) (18, 0.006) (19, 0.005) (20, 0.003) (21, 0.005) (22, 0.005) (23, 0.006) (24, 0.007) (25, 0.007) (26, 0.007) (27, -0.06) (28, -0.435) (29, -0.638) 
    };
    \end{groupplot}
    \end{tikzpicture}

    \end{minipage}
    \hfill
    \begin{minipage}[b]{0.48\linewidth}
        \centering
            \begin{tikzpicture}
    \begin{groupplot}[
    group style={
        group name=my plots,
        group size=1 by 2,
        xlabels at=edge bottom,
        xticklabels at=edge bottom,
        vertical sep=5pt
    },
        width=\linewidth, 
        height=5cm,
        xlabel={Time (minutes)},
        ylabel={Cloudiness},
        xmin=0, xmax=30,
        ymin=0, ymax=1,
        grid=major,
        xtick={0, 5, 10, 15, 20, 25, 30},
        after end axis/.append code={\pgfextra{\vspace{-\baselineskip}}}
    ]
    \nextgroupplot
    \addplot[color=red, mark=star*, mark options={fill=red, scale=0.6}] 
        coordinates {
(0, 0.004) (1, 0.007) (2, 0.242) (3, 0.018) (4, 0.217) (5, 0.341) (6, 0.643) (7, 0.39) (8, 0.927) (9, 0.893) (10, 0.999) (11, 0.815) (12, 0.806) (13, 0.922) (14, 0.955) (15, 0.882) (16, 0.968) (17, 0.88) (18, 0.998) (19, 1) (20, 0.974) (21, 0.917) (22, 1)     };
    
    \addplot[color=blue, mark=square*, mark options={fill=blue, scale=0.6}] 
        coordinates {
(0, 0.004) (1, 0.01) (2, 0.245) (3, 0.063) (4, 0.203) (5, 0.498) (6, 0.52) (7, 0.405) (8, 0.887) (9, 0.978) (10, 0.983) (11, 1) (12, 0.845) (13, 0.477) (14, 0.835) (15, 0.797) (16, 1) (17, 1) (18, 1) (19, 1) (20, 1) (21, 1) (22, 1)   };
    
    \addplot[color=green, mark=triangle*, mark options={fill=green, scale=0.6}] 
        coordinates {
(0, 0.004) (1, 0.004) (2, 0.004) (3, 0.004) (4, 0.004) (5, 0.004) (6, 0.004) (7, 0.004) (8, 0.004) (9, 0.004) (10, 0.004) (11, 0.004) (12, 0.004) (13, 0.004) (14, 0.004) (15, 0.004) (16, 0.004) (17, 0.004) (18, 0.004) (19, 0.004) (20, 0.004) (21, 0.004) (22, 0.004) (23, 0.004) (24, 0.004) (25, 0.004) (26, 0.004) (27, 0.004) (28, 0.004) (29, 0.004) 
    };
    \nextgroupplot[ymin=-1, ymax=1, height=3cm, ylabel={Residuals}, ytick={-1, -0.5, 0, 0.5, 1}, grid=major, xmajorgrids=false]
       \addplot[ycomb, color=orange, mark=*, mark options={fill=orange, scale=0.6}] 
        coordinates {
(0, 0) (1, -0.224) (2, -0.221) (3, 0.384) (4, 0.386) (5, 0.148) (6, 0.309) (7, 0.203) (8, 0.335) (9, -0.103) (10, 0.023) (11, 0.282) (12, 0.137) (13, -0.432) (14, -0.422) (15, -0.064) (16, -0.054) (17, 0) (18, 0) (19, 0) (20, 0) (21, 0) (22, 0)  
    };
    
    \addplot[ycomb,color=purple, mark=square*, mark options={fill=purple, scale=0.6}] 
        coordinates {
(0, 0) (1, -0.005) (2, -0.24) (3, -0.059) (4, -0.199) (5, -0.493) (6, -0.516) (7, -0.4) (8, -0.883) (9, -0.973) (10, -0.978) (11, -0.996) (12, -0.241) (13, -0.473) (14, -0.831) (15, -0.792) (16, -0.996) (17, -0.996) (18, -0.996) (19, -0.996) (20, -0.996) (21, -0.996) (22, -0.996) 
    };
    \end{groupplot}
    \end{tikzpicture}

    \end{minipage}
    \caption{
        The plots show the accuracy for 6 different predictions: Plot 1 (top left) starting at 03:27 on 2020, April 9th, Plot 2 (top right) starting at 02:22 on 2020, April 8th, Plot 3 (middle left) starting at 20:15 on 2023, November 25th, Plot 4 (middle right) starting at 22:50 on 2021, April 26th, Plot 5 (bottom left) starting at 23:51 on 2020, December 31st, Plot 6 (bottom right) starting at 06:16 on 2020, October 18th, . The graphs in the top-part show the \textit{cloudiness} in percent for the \textit{next x minutes}. The cloudiness is a measure for the amount of cloud-pixels in the area of a scheduled observation (a circular region with a radius of 20 pixels around the target) over the total amount of pixels in this area. The green graph (the so-called baseline) represents the cloudiness obtained using the cloud map from the last all-sky image as a reference for the future. The red graph (prediction) shows the cloudiness obtained using the prediction method presented in this paper. The blue graph (ground truth) shows the actual cloudiness measured at that point in time. This is put in as a reference, as this data would not be available in a real-time scenario. The residuals show the difference between the baseline and the prediction (orange) as well as the baseline and the ground truth (red), again, put in as a reference. The predictions in plots 3, 4, and 5 do not have the full prediction dataset, because the observations were made in the area of the sky where the wind was coming from, thus there is not enough cloud data for a longer prediction. Plot 5 shows passing clouds. Plot 6 ends after 22 minutes as the weather control system shut the telescope enclosure because of incoming severe conditions.
    }
    \label{fig:trackresult}
\end{figure*}
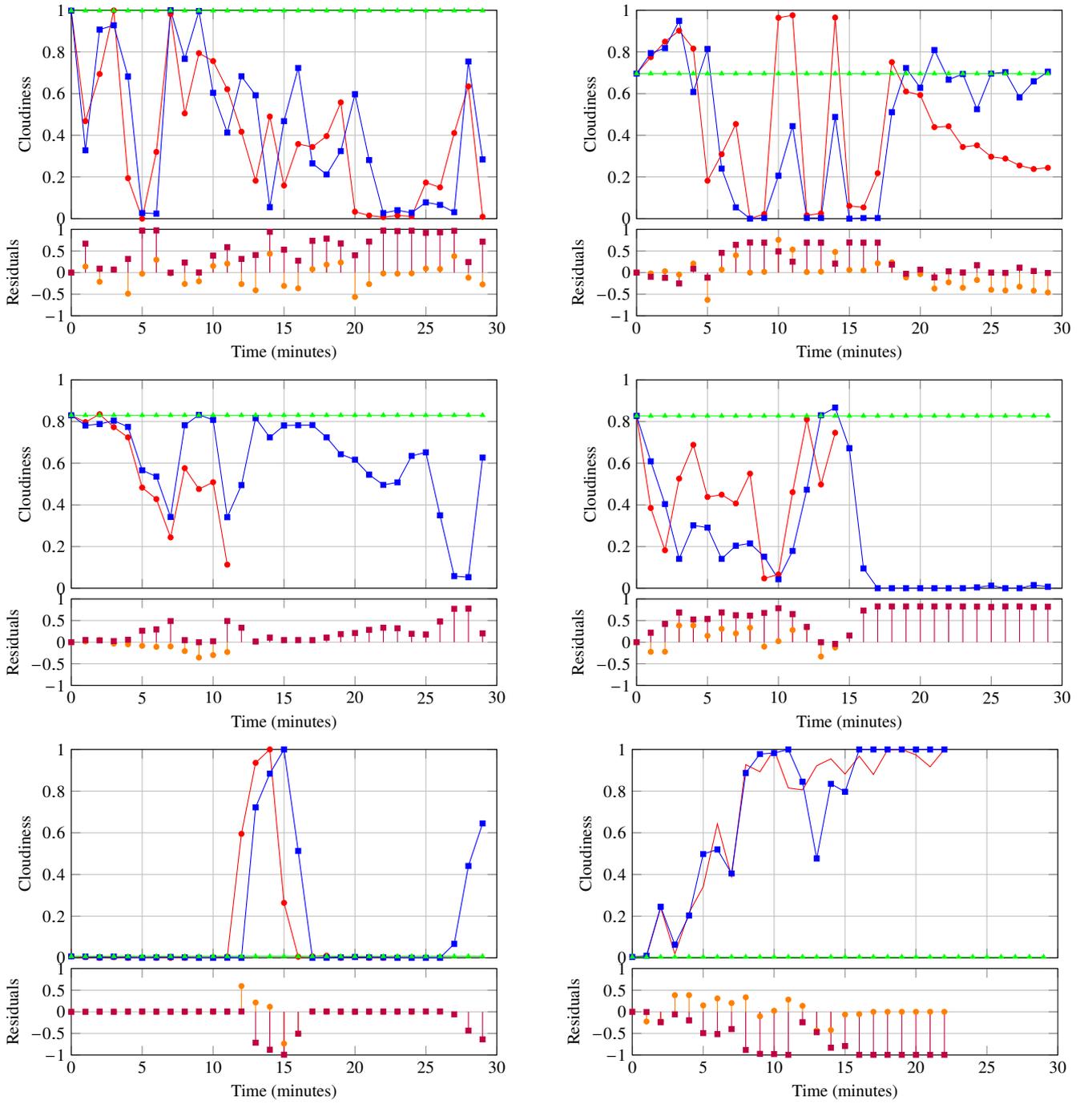


The cloud tracking was tested against historic data. Out of the large dataset of roughly 620,000 images from Skycam A covering the years 2019 to 2022, several nights with cloud cover were selected. The test involved analysing each night, and — if cloud movement was detected — applying the tracking algorithm described in section \ref{sec:cloudmovementprediction}. This was done by simulating the cloud detection system in real-time: images were fed sequentially into the system, which then performed both cloud detection and tracking.

For randomly selected moments during each night, the prediction module was triggered, generating predicted cloud maps for each minute up to 30 minutes into the future. These predictions were then compared to the actual cloud maps at those future time points. Since such ``future'' maps would not be available during live operations, the prediction was also compared against the latest available cloud map (i.e., the starting point of the prediction series), to test whether the prediction provided a measurable improvement.

We selected nights that reflect realistic mixed conditions — clear nights and nights with intermittent or partial cloud cover. The cloud tracking algorithm is not relevant during long overcast periods (in which satellite forecasts suffice and observations are suspended), so the chosen nights represent scenarios where real-time prediction could inform scheduling decisions.

To compare the predicted and actual cloud maps, we extracted a circular region with a radius of 25 pixels around a representative target in each image. In each region, we calculated the ``cloudiness'' parameter, defined as the ratio of cloud-pixels to total pixels. These values were compared for the predicted, actual, and baseline (non-predicted) maps. Figure \ref{fig:trackresult} shows six representative sequences.

These six sequences shown in Figure \ref{fig:trackresult} were selected to illustrate a range of cloud behaviours and prediction outcomes. They are not chosen for peak performance but rather to reflect typical scenarios the system encounters. We have excluded rare edge cases where prediction accuracy extended beyond 30 minutes, as these were atypical and not statistically representative.

The main limiting factor is the distortion near the horizon due to the fisheye projection. Pixels close to the horizon correspond to large, distorted regions of the sky, and incoming clouds in these areas cannot be reliably predicted. The effective forecast horizon is therefore strongly influenced by wind speed and direction: slower-moving clouds can be tracked longer, while fast-approaching clouds from off-frame regions limit the prediction range.

The expected scheduling strategy for the New Robotic Telescope (NRT) is a look-ahead approach, with a planning horizon of approximately 15–20 minutes. As NRT will be required to respond to transient alerts (e.g., from the Vera Rubin Observatory, ZTF, LIGO, CTA), this prediction window aligns well with the typical duration of high-priority observations (usually around 10 minutes). Reliable short-term sky forecasts can thus play a valuable role in optimizing telescope scheduling, particularly under dynamic or marginal weather conditions.

\begin{figure}
\centering
\begin{tikzpicture}
\begin{axis}[
    width=\columnwidth, height=0.6\columnwidth,
    xlabel={Time since start [minutes]},
    ylabel={Agreement},
    grid=both,
    xmin=0, xmax=25, 
    legend style={at={(0.5,1.05)},anchor=south},
    ymin=0, ymax=1,
    cycle list name=color list,
    legend columns=3,
    legend cell align={left},
]

\addplot table[x index=0, y index=3, col sep=comma] {cloudcovernorm.csv};
\addlegendentry{Avg Agreement}

\addplot table[x index=0, y index=1, col sep=comma] {cloudcovernorm.csv};
\addlegendentry{Min Agreement}

\addplot table[x index=0, y index=2, col sep=comma] {cloudcovernorm.csv};
\addlegendentry{Max Agreement}

\addplot [
    name path=upper,
    draw=none,
] table[x index=0, y index=2, col sep=comma] {cloudcovernorm.csv};

\addplot [
    name path=lower,
    draw=none,
] table[x index=0, y index=1, col sep=comma] {cloudcovernorm.csv};

\addplot [
    fill=gray!10,
] fill between[of=upper and lower];

\end{axis}
\end{tikzpicture}
\caption{Agreement between predicted and actual cloud maps over time, averaged across 100 sequences. The red line shows the average agreement, while the shaded area spans the minimum and maximum values observed at each time step. Agreement is defined as the fraction of correctly matched cloud pixels in the region of interest.}
\label{fig:agreementpredict}
\end{figure}
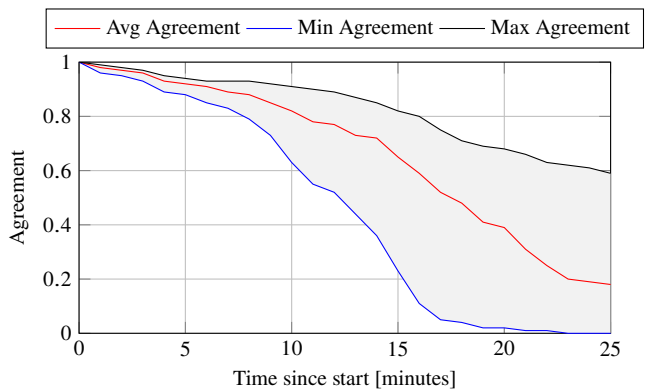

As can be seen in the data (see figure \ref{fig:trackresult}), a prediction up to 30 minutes is usable. During the analysis of the historic data of the Liverpool Telescope it became clear that most groups - once picked by the scheduler and started by the Robotic Control System (see \cite{fraser2004}) - are finished after 15 to 20 minutes matching well with the prediction horizon of the tracking and prediction algorithm presented in this paper.

To compare the two maps, a small circular region with radius of 25 pixels around the target was extracted from the `prediction cloud map', the actual cloud map at the point in time the prediction represents and the cloud map at the starting point of the prediction (the latest available one in a real scenario). Then, the cloudiness parameter was calculated (the ratio of cloud pixels over total pixels) in each region and plotted in figure \ref{fig:trackresult}.

The results show that predictions up until 15 minutes into the future can be made with a high accuracy. After this time, the accuracy drops. Predictions for more than 30 minutes are unreliable. Depending on the location of the target of the sky and the wind direction, the prediction times can also be significantly shorter (e.g. if the wind is coming from the east and a target is in the eastern part of the sky, the data needed to do a prediction of 15 minutes and more would lie outside the image bounds).

The main limiting factor is that only little knowledge of incoming clouds is available in the images. Due to the fish-eye effect, a pixel close to the horizon spans a large area of the sky with an extremely high distortion compared to pixels in the centre of the image. This causes the incoming clouds to be severely distorted in a way that can not be fully accounted for in the algorithm. It is thus not possible to reconstruct the cloud pattern after a certain time. This time is closely correlated to the wind speed in the cloud layer. In low wind speeds the clouds move slower and predictions can be made up to 20 minutes more. Two cases were found (but not included in the data used for the graph in figure \ref{fig:trackresult}) were predictions up to 35 minutes still provided usable results, but as these were rare, hand-picked events, they were excluded. A study is in preparation to see if a correlation between the wind-speed at ground level and the cloud speed can be determined. For a more detailed explanation of the atmospheric properties of La Palma, \citep{varela2008} provides a good overview. 

To provide a statistically robust evaluation of the cloud tracking and prediction algorithm, we analysed 100 cloud sequences extracted from historical Skycam A data between 2020 and 2022. Nights were pre-filtered to exclude fully clear or fully overcast conditions using a threshold on global cloudiness, and from the remaining mixed-condition nights, sequences of at least 30 minutes with visible cloud motion were randomly selected. For each sequence, the system generated predicted cloud maps up to 25 minutes into the future. These were compared against the actual cloud maps at each time step using a circular region of interest (ROI) with 25-pixel radius, randomly placed within the altitude range of $35^\circ$ to $80^\circ$ to minimise distortion effects near the horizon and zenith. The agreement metric, defined as the fraction of matching cloud pixels in the ROI, was used to quantify prediction accuracy. The results, shown in Figure~\ref{fig:agreementpredict}, present average, minimum, and maximum agreement values across all 100 sequences. Agreement remains high during the first 10–15 minutes and decreases gradually beyond that, depending on cloud structure and dynamics. A second, independently selected test set confirmed that 100 sequences are sufficient to yield stable statistics (agreement deviation <3\%). The six example sequences in Figure~\ref{fig:trackresult} were included to illustrate \textit{typical} visual behaviour and were not selected based on performance quality.

All evaluation was done using the all-sky camera sequence itself as ground truth. We did not use Skycam T for this purpose because its small and variable field of view, dictated by the telescope’s pointing, prevents consistent tracking of cloud structures over time.

The planned scheduling strategy for the New Robotic Telescope is a look-ahead scheduling strategy with a time horizon of about 15 to 20 minutes into the future. As the NRT will be used to react to alerts from optical surveys as the Vera Rubin Observatory or the Zwicky Transient Facility as well as other surveys like the LIGO Gravitational Wave Observatory or the Cherenkov Telescope Array. The NRT will thus have to process many different targets with short exposures (around 10 minutes of exposure time) so a prediction of the sky conditions over the next 10 to 30 minutes will be a useful contribution to the scheduling decision, especially in mixed conditions.

\section{Conclusion}

In this paper a novel set of algorithms for cloud detection, tracking and the prediction of their future location were presented. The cloud detection algorithm uses difference images taken on regular intervals with an all-sky camera and uses binarization techniques to detect clouds in these all-sky images.  

The algorithm  is fast and - with the use of libraries as Astrometry.net \citep{lang2010} - does only need little calibration, depending on the exposure settings of the all-sky camera. Re-calibration is only necessary if the camera hardware itself or the orientation is changed. It can be plugged into the data feed of an all-sky camera and provide cloud patterns for Right Ascension and Declination coordinates provided to a Web-API. The system should ideally be positioned so that moving telescope parts are not visible (as it is the case at the LT) to avoid either masking a rather large area or misclassify the moving structure as cloud.

The robustness of the algorithm was tested against \citep{adam2017} and provided comparable results in dark nights and outperformed the other system in bright nights. 

The main limiting factor of this algorithm is the binary decision if a pixel represents either `cloud' or `clear sky' and is not able to determine the opacity of the cloud e.g. how much fainter a star would be through a thin cloud. A future version will implement the photometric approach of \citep{adam2017} in combination with the presented cloud detection by measuring visible stars in the area of a cloud pattern.

The cloud tracking and detection system sits on top of the cloud detection system and analyses a time-series of clouds to detect individual clouds and track their movement through the image plane to determine a common tracking vector. This tracking vector is then used to predict future positions of the cloud pixels and generated predicted cloud patterns. Up to 15 minutes in the future (with longer prediction times possible with lower wind speeds in the cloud layer), predictions with acceptable accuracy can be made. This is a sufficient amount of time for the look-ahead scheduler planned for the New Robotic Telescope.

The system makes use of widely available hardware to provide real-time results of the cloud patterns and the predictions and provides a calibration-free setup as well as a Web-API for easy access to the data. 

The results showed that clouds can be detected with a very high precision with a false positive rate of around $1\%$ (in areas with an angular distance $>30\%$ of the moon) and a false negative rate of $1\%$, except in dark nights with slow-moving clouds. The cloud prediction can be performed with an accuracy average of $65\%$ for a prediction of 15 minutes into the future. 

\section*{Acknowledgements}
The Liverpool Telescope is operated on the island of La Palma by Liverpool John Moores University in the Spanish Observatorio del Roque de los Muchachos of the Instituto de Astrofisica de Canarias with financial support from the UK Science and Technology Facilities Council. The telescope and Chris M. Copperwheat receive funding from UKRI grant number ST/X005933/1. Chris M. Copperwheat and Helen E. Jermak receive funding from UKRI grant number ST/W001934/1. We thank the anonymous referee for their useful comments which led to some significant improvements to this paper.

\section*{Data Availability}
\label{sec:data}

All data used for this paper is available as FITS files at \url{https://telescope.livjm.ac.uk} in the Skycam Archive. The source code is available on request from the authors and will be made public once the underlying project is finished.



\bibliographystyle{rasti}
\bibliography{example.bib} 








\bsp	
\label{lastpage}
\end{document}